\documentclass[fp,twocolumn]{jpsj3}
\usepackage{graphicx,amsmath,amssymb,bm}
\usepackage{braket}

\allowdisplaybreaks 

\usepackage{color}
\definecolor{Green}{rgb}{0,0.7,0}

\newcommand{\e}{ {\rm e}}

\newcommand{\BETS}{ $\alpha$-(BETS)$_2$I$_3$}
\newcommand{\bk}{ \bm{k}}
\newcommand{\bG}{ \bm{G}}
\newcommand{\bkD}{ \bm{k}_{\rm D}}
\newcommand{\eD}{ \epsilon_{\rm D}}
\newcommand{\ep}{ \epsilon }
\newcommand{\hn}{ \hat{n}}
\newcommand{\tV}{{V}}
\newcommand{\kx}{k_x}
\newcommand{\ky}{k_y}
\newcommand{\g}{ \gamma }
\newcommand{\bq}{ \bm{q}}
\newcommand{\bX}{\bar{X}}
\newcommand{\bY}{\bar{Y}}

\title{
Electric and Magnetic Responses 
 of Two-dimensional Dirac Electrons in  
Organic Conductor $\alpha$-(BETS)$_2$I$_3$ }
\author{
Yoshikazu Suzumura\thanks{E-mail: suzumura.yoshikazu@d.mbox.nagoya-u.ac.jp}
$^{1}$
 and Takao Tsumuraya\thanks{E-mail:  tsumu@kumamoto-u.ac.jp}
$^{2}$
}
\inst{
$^1$Department of Physics, Nagoya University,  Nagoya 464-8602, Japan \\
$^2$
Priority Organization for Innovation and Excellence, Kumamoto University, Kumamoto 860-8555, Japan
}
\recdate{7/20/2021, Accepted 10/25/2021}
\abst{
Effect of spin-orbit coupling (SOC) on Dirac electrons in the organic conductor $\alpha$-(BETS)$_2$I$_3$
 [BETS = bis(ethylenedithio)tetraselenafulvalene]  has been examined by calculating electric conductivity and spin magnetic 
 susceptibility. A tight-binding (TB) model with real and imaginary transfer energies is derived using first-principles density-functional theory method. The conductivity without the SOC depends on both  anisotropies of the velocity of the Dirac cone  and the tiling of the cone.
Such conductivity is suppressed by the SOC, which gives rise to the imaginary part of the transfer energy.
 Due to the  SOC,  we find at low temperatures that the reduction of the conductivity becomes large and that the anisotropy of the conductivity is reduced. A nearly constant conductivity at high temperatures is obtained by an  electron--phonon (e--p) scattering. 
Further, the property of the Dirac cone is examined for the spin susceptibility, which is mainly determined by the density of states (DOS).
The result is compared with the case of the organic conductor $\alpha$-(BEDT-TTF)$_2$I$_3$ [BEDT-TTF=bis(ethylenedithio)tetrathiafulvalene], which provides the Dirac cone without the SOC.
 The relevance to experiments is discussed. 
  }

\begin{document}
\maketitle
\section{Introduction} 
Since the discovery in the graphene\cite{Novoselov2005_Nature438}, massless Dirac fermion has been studied extensively. 
Especially  the Dirac electron  in an organic conductor, $\alpha$-(BEDT-TTF)$_2$I$_3$ [BEDT-TTF=bis(ethylenedithio)tetrathiafulvalene], was found as a bulk system.\cite{Kobayashi2004,Katayama2006_JPSJ75,Kajita_JPSJ2014}
The two-dimensional  Dirac cone provides the density of states (DOS) vanishing  linearly at the Fermi energy and  a zero-gap state (ZGS).
The energy  band  was calculated using a tight-binding (TB) model, where  transfer energies are estimated from the extended H\"uckel method.\cite{Kondo2005} 
Such a Dirac cone  was verified by a first-principles density functional theory (DFT) calculation.\cite{Kino2006}

The ZGS  has been studied in  some organic conductors with  
 isostructural salts,
$\alpha$-D$_2$I$_3$ (D = ET, STF, and  BETS), where 
ET = BEDT-TTF, 
 STF = bis(ethylenedithio)diselenadithiafuluvalene), 
 and 
BETS =  bis(ethylenedithio)tetraselenafulvalene.
These salts display  an energy band  with a Dirac cone,
\cite{Katayama2006_JPSJ75,Kondo2009,Morinari2014,Naito2020} 
 and   the resistivity at high temperatures
 shows a nearly constant behavior.
\cite{Inokuchi1995_BCSJ68,Kajita1992,Tajima2000,Tajima2002,Tajima2007,
Liu2016
}
Such a constant resistivity behavior  at high temperatures 
 suggests a common feature 
    of  Dirac electrons in organic conductors.
  In contrast,
 the  resistivity at low temperatures  shows a different behavior 
     depending on the salts. 
For $\alpha$-(ET)$_2$I$_3$, 
 the insulating state is obtained by  the charge ordering (CO),   
 where A and A' molecules in the unit cell  become  inequivalent 
 due to  
  breaking  the inversion symmetry.
\cite{Takano2001,Wojciechowskii03,Kakiuchi2007} 
Under high pressures, the CO is absent. The resistivity  
   shows only a slight enhancement with a minimum,\cite{Tajima2007} 
 and  the equivalence of  A and A' due to the inversion symmetry 
 is shown  
from the 
 spin susceptibility.\cite{Katayama_EPJ,Takahashi2010,Hirata2016} 
For $\alpha$-(BETS)$_2$I$_3$,  the CO is absent,  and 
the resistivity shows  an enhancement but not  an insulating state 
 at ambient pressure, and 
   an almost constant  behavior under pressure. 

The physical properties of Dirac electrons have been studied in terms of 
 an effective Hamiltonian of the two-band model.
\cite{Kobayashi2007,Goerbig2008,Kobayashi2008}
The conductivity with  zero doping 
has been studied theoretically using a two-band model 
 with  a simple Dirac cone.
 The static conductivity 
at absolute zero temperature  remains finite with a universal value, i.e., independent of the magnitude of impurity scattering  
 owing to a quantum effect.\cite{Ando1998} 
The effect of   Dirac cone tilting  
  shows  
 the anisotropic conductivity   
 and  the deviation of the current from an applied electric field.
\cite{Suzumura_JPSJ_2014}
At finite temperatures, on the other hand, 
the conductivity  depends on the magnitude 
 of the impurity scattering, $\Gamma$, which is proportional to the inverse of 
 the life time by the disorder. 
With increasing $T$, the conductivity remains unchanged 
for $T \ll \Gamma$,
 whereas it increases for $\Gamma \ll T$.\cite{Neto2006} 
Noting that  $\Gamma \sim$ 0.0003 eV for organic conductors,\cite{Kajita_JPSJ2014}
 a  monotonic increase in 
the conductivity at finite temperature $T > 0.0005$ eV
 is expected.
However,  the measured conductivity (or resistivity) of the above organic conductor 
 shows  an almost constant behavior at high temperatures. 
 To comprehend such an exotic  phenomenon, 
  the  acoustic phonon scatterings have been  proposed 
  as  a possible mechanism, which was studied 
  using  a simple two-band model of the Dirac cone  without tilting.
~\cite{Suzumura_PRB_2018}
 Such  a  mechanism  reasonably explains   the conductivity
  in $\alpha$-(ET)$_2$I$_3$,
 described by the TB model.~
\cite{Suzumura2021_JPSJ}
However, electric conductivity for $\alpha$-(BETS)$_2$I$_3$
 has yet to be clarified theoretically. 

 The selenium-substituted analog $\alpha$-(BETS)$_2$I$_3$ has recently  attracted  attention as a possible Dirac electron at ambient pressure.
The temperature crossover from the metal to insulating behavior around 50 K ($T_{MI}$) is lower than the CO transition temperature 
in $\alpha$-(ET)$_2$I$_3$.\cite{Inokuchi1995_BCSJ68}
To understand the origin of the increased resistivity at low temperatures, several groups studied whether the presence or absence of the  CO transition at ambient pressure.
At high temperatures, the spin susceptibility is similar between $\alpha$-(ET)$_2$I$_3$ 
and $\alpha$-(BETS)$_2$I$_3$.\cite{Takahashi2011_JPSJ80} 
However, the NMR suggests the inversion symmetry, which would indicate the absence of the CO in $\alpha$-(BETS)$_2$I$_3$.\cite{Shimamoto2014}
Structural analysis based on x-ray diffraction also suggests no breaking the inversion symmetry around the $T_{MI}$~\cite{Kitou2020}.

First-principles calculation for the low-temperature structure reveals a pair of anisotropic Dirac cones at a general $\bk$-point, when the spin-orbit coupling (SOC) effect is ignored~\cite{Kitou2020}. 
The Dirac cone band structure is robust, which is different from a previous DFT band structure for the 0.7~GPa structure, where Dirac point and electron-hole pockets coexist~\cite{Alemany2012}.

In contrast, when we consider the SOC effect, an indirect gap of $\sim$2~meV is opened at the Dirac points.  
The band gap size is generally consistent with the $T_{MI}$, since (semi) local density approximation in DFT slightly underestimates the experimental energy gap. The calculated $\mathbb{Z}_2$ topological invariant suggests the system is a weak topological insulator~\cite{Kitou2020}.  
Shubnikov-de Haas oscillation measurement verified the absence of massless Dirac electrons at ambient pressure and low temperature, 
and the Dirac fermion phase appears under pressure~\cite{Tajima2021PRB}. 
Since the width of linear band dispersion is wider than the band gap, 
 this system can exhibit the behavior of a Dirac electron at finite temperature.
Moreover, it is not clear how the velocity anisotropy of the Dirac cone affects the electrical conductivity.

To derive a TB model from DFT band dispersions is crucial to comprehend the Dirac electrons  properly.\cite{Konschuh2010} 
However, efficient methods for extracting effective TB models, 
 including the SOC, have not been fully established for molecular solids.\cite{Sanvino2017, Winter2017} 
In our previous work, it is found that the delocalized character of Se $p$ orbitals constrains the eigenvalues close to the Dirac points in a quite-narrow energy window, compared with the electronic state of $\alpha$-(ET)$_2$I$_3$~\cite{EPJB2020}. Therefore, the result of fitting to the DFT bands indicates that the number of relevant transfer integrals is significant.~\cite{EPJB2020}
To develop a reliable low-energy model Hamiltonian with a moderate number of transfer energies, we introduce site-potentials, which reasonably reproduce the spectrum of the DFT eigenvalues at several time-reversal invariant momenta (TRIM), and propose a precise TB model for the insulating state in $\alpha$-(BETS)$_2$I$_3$.

In this paper, we study the effect of the SOC on  the anisotropic conductivity 
  using the TB model of $\alpha$-(BETS)$_2$I$_3$,\cite{EPJB2020}
 which contains  both real and imaginary parts
   in  the transfer energy.
 By using such a TB model,
we clarify the origin of the insulating behavior at low temperatures. 
 It is also  shown that the presence of  acoustic phonons gives rise to 
 the  conductivity being nearly constant at high temperatures.
The paper is organized as follows.
 In Sect. 2, the model and formulation are given 
  for $\alpha$-(BETS)$_2$I$_3$.
In Sect. 3, after examining the chemical potential and density of states (DOS),
 the conductivity and spin susceptibility are calculated, 
 and the characteristics are demonstrated by  comparing 
with those of $\alpha$-(ET)$_2$I$_3$.
~\cite{Suzumura2021_JPSJ}
 Section 4 is devoted to a summary and discussion of the relevance to experiments.

\section{Model and Formulation}
\begin{figure}
\vspace{0.5cm}
  \centering
\includegraphics[width=0.8\linewidth]{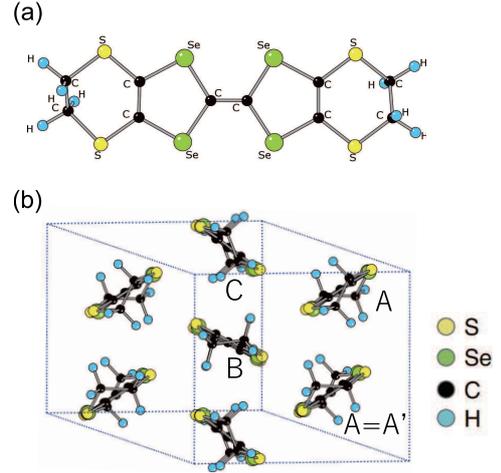}
     \caption{(Color online)
(a)  Molecular structure of 
   BETS (=  bis(ethylenedithio)tetraselenafulvalene), 
 which has a glide symmetry  to the center of 
  the    C=C bond. 
(b) Crystal structure of $\alpha$-(BETS)$_2$I$_3$ 
 with  four molecules of  
  A, A', B, and C in the unit cell forming  a square lattice. 
 The inversion center exists 
 at the middle of A and A', which are equivalent molecules. 
 Transfer energies  taken  for    
  nearest neighbor (NN) 
 and next-nearest neighbor (NNN) sites 
   are estimated using  
 the Wannier function with the center 
  set at the center of the C=C double bond of each molecule. 
}
\label{fig1}
\end{figure}

We consider a two-dimensional Dirac electron system,
 which is given by 
\begin{equation}
H_{\rm total}= H_0 + H_{1} + H_{\rm p} +  H_{\rm e-p} +H_{\rm imp} \; . 
\label{eq:H}
\end{equation}
 $H_0$ describes a TB model of 
 the  organic conductor \BETS\;  consisting of four molecules per unit cell 
(Fig.~\ref{fig1}). 
 $H_{1}$ represents a site potential,\cite{EPJB2020}
 which originates from the Hartree term of the Coulomb interaction. 
 $H_{\rm p}$ and 
 $H_{\rm e-p}$ denote  an acoustic phonon and 
an  electron-phonon (e--p) interaction, respectively. 
$H_{\rm imp}$ is the impurity potential. 
The unit of the energy is taken as eV. 
 The lattice constant is taken as unity.
\subsection{Energy band}
First, we  calculate the energy band for $H=H_0+H_1$ 
and the associated quantities.
A TB model, $H_0$, is expressed as 
\begin{eqnarray}
H_0 &=& \sum_{i,j = 1}^N \sum_{\alpha, \beta = 1}^4 \sum_{s, s'= \pm}
 t_{i,j; \alpha s,\beta s'} a^{\dagger}_{i,\alpha,s} a_{j, \beta,s'} 
                          \nonumber \\
&=& \sum_{\bk}  \sum_{\alpha, \beta = 1}^4  
\sum_{s, s'= \pm}
 t_{\alpha s, \beta s'}(\bk)  a^{\dagger}_{\alpha s}(\bk) a_{\beta s'}(\bk) 
\; , 
\label{eq:Hij}
\end{eqnarray}
where   $a^{\dagger}_{i, \alpha, s}$ denotes a creation operator 
 of an electron 
 of molecule $\alpha$ 
 [= A(1), A'(2), B(3), and C(4)] with spin $s=\pm$ in the unit cell 
  at  the $i$-th lattice site.
$s=+$ and $s=-$ denote $\uparrow$ and $\downarrow$  spins. 
 $N$ is the total number of square lattice sites and 
 $t_{i,j; \alpha s,\beta s'}$ are the transfer energies 
for  the nearest  neighbor (NN) and next-nearest 
 neighbor (NNN) sites.~\cite{EPJB2020} 
A Fourier transform for the operator $a_{j,\alpha,s}$ 
 is given by   
 $a_{j,\alpha, s} = 1/N^{1/2} \sum_{\bk} a_{\alpha s}(\bk) \exp[ i \bk \cdot \bm{r}_j]$.
 The wave vector  $\bk = (k_x,k_y)$ 
 is taken within  $\bG$, which  
 denotes  a reciprocal lattice vector of the square lattice. 
 The quantity $\bG$/2 
  corresponds  to the vector of the time-reversal invariant momentum (TRIM).
The quantity $H_1$ corresponds to   a site potential, 
 $V_{\alpha}$,  acting on the electron at 
 the $\alpha$ site, where 
  $V_{\rm A} = V_{\rm A'}$ due to an inversion symmetry 
  around the middle point between A and A' molecules in Fig. \ref{fig1}.  
 The Hamiltonian $H_1$ is obtained as (Appendix A)
\begin{eqnarray}
H_1 & = &\sum_{\alpha} (V_{\alpha} - V_{\rm A})\hn_{\alpha} 
                   \nonumber \\
    & = & \Delta \tV_{\rm B} \hn_{\rm B} + \Delta \tV_{\rm C} \hn_{\rm C} \; , 
\label{eq:H1}
\end{eqnarray}
 where $ \Delta \tV_{\alpha}$ denotes a  potential measured from that 
 of the A-site  
 and  $\hn_{\alpha} = \sum_{\bk} \sum_s a^{\dagger}_{\alpha  s}(\bk) a_{\alpha s}(\bk)$.  
From Eqs.~(\ref{eq:Hij}) and (\ref{eq:H1}),
 $H$ is written as\cite{Katayama_EPJ} 
\begin{eqnarray}
  H = \sum_{\bk} \sum_{\alpha, \beta} \sum_{s, s'} 
      a^{\dagger}_{\alpha s}(\bk) 
        h_{\alpha s,\beta s'} a_{\beta s'}(\bk)\; , 
\label{eq:H_total}
\end{eqnarray}
where $h_{\alpha s,\beta s'}$ denotes the matrix element (Appendix A). 
Noting that eigenvalues are degenerate with respect to spin, 
 Eq.~(\ref{eq:H_total}) is diagonalized as 
\begin{subequations}
\begin{eqnarray}
\label{eq:eq6a}
 H = \sum_{\bk} \sum_{\g} \sum_{s}
  c_{\g s}^{\dagger}(\bk) E_\g(\bk) c_{\g s}(\bk) \; ,
\end{eqnarray}
where $E_1(\bk) > E_2(\bk) > E_3(\bk) > E_4(\bk)$  and 
\begin{eqnarray}
& & \sum_{\beta} \sum_{s'}h_{\alpha s,\beta s'}(\bk)
           d_{\beta s' \g }(\bk)
   = E_{\g}(\bk) d_{\alpha s \g } (\bk)  \; , 
        \nonumber \\
\label{eq:eq6b}
& & c_{\g s}(\bk) = 
   \sum_{\alpha} d_{\alpha s \g}(\bk)  a_{\alpha s}(\bk) \; .
\label{eq:eq6c}
\end{eqnarray}
\end{subequations}
The Dirac point ($\bkD$) is calculated  from  
\begin{eqnarray}
\label{eq:ZGS}
E_1(\bkD) = E_2(\bkD)= \eD \; .
\end{eqnarray}
 The ZGS is obtained  when 
 $\eD$ becomes equal to  the chemical potential at $T = 0$. 

From $E_\g$, the local density  $n_{\alpha}$ 
including  both spin $\uparrow$ and $\downarrow$ 
   is calculated as   
\begin{eqnarray}
 n_{\alpha} & = &\frac{1}{N} \sum_{\bk} \sum_{s}
     \left< \hn_{\alpha}(\bk) \right>_{H} \nonumber \\ 
 & = &
 \frac{1}{N} \sum_{\bk}\sum_{\g} \sum_{s}
  d^*_{\alpha s \g}(\bk) d_{\alpha s \g}(\bk)f(E_\g(\bk)-\mu)
 \; , \nonumber \\
\label{eq:local_charge}
 \end{eqnarray}  
 which is determined self-consistently. 
 $n_{\rm A} = n_{\rm A'}$  owing  to transfer energies  
 being symmetric  as for the inversion center   between   A and A'
 in Fig.~\ref{fig1}.
In Eq.~(\ref{eq:local_charge}), $f(\ep)= 1/(\exp[\ep/T]+1)$ with $T$ being temperature in the unit of eV 
 and $k_{\rm B }=1$.
The chemical potential $\mu$ is determined 
 from the three-quarter-filled condition, which is given by 
\begin{eqnarray}
  \frac{1}{N} \sum_{\bk} \sum_{\gamma}  f(E_{\gamma}(\bk)-\mu)=
 \int_{-\infty}^{\infty} {\rm d} \omega D(\omega) f(\omega) =  3 \; ,  
  \label{eq:mu}
\end{eqnarray}
where  
\begin{eqnarray}
D(\omega) &=& \frac{1}{N} \sum_{\bk} \sum_{\gamma}
 \delta (\omega - E_{\gamma}(\bk)) \; .
  \label{eq:dos}
\end{eqnarray}
$D(\omega)$ denotes DOS per spin and per unit cell, 
 which satisfies  $\int {\rm d} \omega D(\omega) = 4$.
Note that $ n_{\rm A} + n_{\rm A'} + n_{\rm B} + n_{\rm C} =6$
from Eq.~(\ref{eq:mu}).
We use $\mu (T)$ at finite $T$ and $\mu$ = $\mu(0)$ at $T$=0. 

\subsection{Conductivity}
By using the component of the wave function $d_{\alpha \gamma}$ 
 in Eq.~(\ref{eq:eq6b}), we calculate 
 the conductivity   per spin  
   as\cite{Katayama2006_cond}  
\begin{eqnarray}
\sigma_{\nu \nu'}(T) &=&  
  \frac{e^2 }{\pi \hbar N} 
  \sum_{\bk} \sum_{\gamma, \gamma',s} 
  v^\nu_{\gamma \gamma's}(\bk)^* 
  v^{\nu'}_{\gamma' \gamma s}(\bk) \nonumber \\
& &  \int_{- \infty}^{\infty} d \ep 
   \left( - \frac{\partial f(\ep) }{\partial \ep} \right)
    \nonumber \\
  \times & &\frac{\Gamma_\g}{(\ep - \xi_{\bk \gamma})^2 + \Gamma_\g^2} \times 
 \frac{\Gamma_{\g'}}{(\ep - \xi_{\bk \gamma'})^2 +  \Gamma_{\g'}^2}
  \; ,  \nonumber \\
  \label{eq:sigma}
\\
  v^{\nu}_{\gamma \gamma's}(\bk)& = & \sum_{\alpha \beta}
 d_{\alpha \gamma s}(\bk)^* 
   \frac{\partial h_{\alpha s, \beta s}}{\partial k_{\nu}}
 d_{\beta \gamma's}(\bk) \; ,
  \label{eq:v}
\end{eqnarray}
 where $\xi_{\bk \gamma} = E_{\g}(\bk) - \mu$, 
$\nu = x$ and $y$.
 $h = 2 \pi \hbar$ and $e$ denote  Planck's constant and electric charge, 
 respectively.  
 The quantity $\Gamma_\g$ indicates 
 the damping of the electron of the $\g$ band given by 
\begin{eqnarray}
\Gamma_{\g}  = \Gamma + \Gamma_{\rm ph}^{\g} \; ,
 \label{eq:g}
\end{eqnarray}
where the first term comes from the impurity scattering 
and the second term corresponding to  the phonon scattering 
 is given by~\cite{Suzumura_PRB_2018} (Appendix B)
\begin{subequations}
\begin{eqnarray}
  \Gamma_{\rm ph}^\g &=&  C_0R \times T|\xi_{\bk \g s}|
  \; ,
 \label{eq:eq16a}
        \\ 
R &=& \frac{\lambda}{ \lambda_0}
 \; ,  
 \label{eq:eq16b} 
\end{eqnarray}
 \end{subequations}
with  $C_0 = 25\lambda_0/(2\pi v^2)$ with freedom of spin and valley.  
 For $v \simeq 0.05$  and $\lambda_0/2\pi v = 0.1 $,
 we obtain $C_0 \simeq$ 50 (eV)$^{-1}$.
$R$  denotes a normalized e--p coupling constant.

Here we note   the damping by the Coulomb interaction, which can be 
 calculated in a way similar to  Eq.~(\ref{eq:self_energy}).
 Although the Coulomb interaction, especially the forward scattering, 
has a  significant  effect of suppressing  the spin susceptibility 
of $\alpha$-(ET)$_2$I$_3$~\cite{Hirata2016}  due to 
the unscreening of the interaction   for the undoped  Dirac cone,
\cite{Kotov2012}
the  effect for the case of  $\alpha$-(BETS)$_2$I$_3$ 
 is considered to be small  
   from the behavior of the susceptibility as shown at the end of Sect. 3.  
Furthermore, since  the Coulomb interaction is an internal 
force,  such an interaction is usually  ignored  for the conductivity.

In the following, we denote $\sigma_x$, $\sigma_y$, and $\sigma_{xy}$ 
instead of $\sigma_{xx}(T)$, $\sigma_{yy}(T)$, and $\sigma_{xy}(T)$ 
 for simplicity. 
In terms of  
  $\sigma_x$,  
 $\sigma_y$, and  
 $\sigma_{x y}$,  
 the current $(j_x,j_y)$ obtained from a response to 
 an external electric field 
 $(E_x,E_y)$  is written as 
\begin{eqnarray}
\begin{pmatrix}
  j_x  \\
  j_y
\end{pmatrix}  
=
\begin{pmatrix}
  \sigma_x  & \sigma_{xy}  \\
  \sigma_{xy}  & \sigma_y 
\end{pmatrix}  
\begin{pmatrix}
  E_x  \\
  E_y
\end{pmatrix} \; . 
 \label{eq:eq19}
\end{eqnarray}
The principal axis of the Dirac cone has an angle $\phi$ 
measured from the $k_y$ axis, where $-\pi/2 < \phi < \pi/2$.
When we denote the current and  electric field in this
   axis direction as $j_x'$ and $E_x'$, 
we obtain 
\begin{subequations}
\begin{eqnarray}
\begin{pmatrix}
  j_x'  \\
  j_y'
\end{pmatrix}  
=
\begin{pmatrix}
  \sigma_-  & 0  \\
  0  & \sigma_+
\end{pmatrix}  
\begin{pmatrix}
  E_x'  \\
  E_y'
\end{pmatrix} \; , 
 \label{eq:eq20a}
\end{eqnarray}
where 
\begin{eqnarray}
\begin{pmatrix}
  j_x'  \\
  j_y'
\end{pmatrix}  
=
\begin{pmatrix}
  \cos \phi & \sin \phi  \\
  - \sin \phi & \cos \phi
\end{pmatrix}  
\begin{pmatrix}
  j_x  \\
  j_y
\end{pmatrix} \; , 
 \label{eq:eq20b}
\end{eqnarray}
%
\begin{eqnarray}
\begin{pmatrix}
  E_x'  \\
  E_y'
\end{pmatrix}  
=
\begin{pmatrix}
  \cos \phi & \sin \phi  \\
  - \sin \phi & \cos \phi
\end{pmatrix}  
\begin{pmatrix}
  E_x  \\
  E_y
\end{pmatrix} \; . 
 \label{eq:eq20c}
\end{eqnarray}
\end{subequations}
Quantities  $\phi$, $\sigma_-$, and $\sigma_+$  are 
 obtained as  
\begin{subequations}   
\begin{eqnarray}
\tan 2 \phi &=& \frac{2\sigma_{xy}}{\sigma_x-\sigma_y} \; ,
    \label{eq:eq21a}
          \\
 \sigma_{-} &=& 
 \frac{1}{2}[\sigma_x+\sigma_y - \sqrt{(\sigma_x-\sigma_y)^2+ 4\sigma_{xy}^2}] \; , 
            \nonumber \\
         \label{eq:21b} 
    \\
  \sigma_{+} & = &
     \frac{1}{2}[\sigma_x+\sigma_y + \sqrt{(\sigma_x-\sigma_y)^2+ 4\sigma_{xy}^2}] \;. 
       \nonumber \\
       \label{eq:eq21c}
\end{eqnarray}
\end{subequations}   
Note that $\sigma_{xy}$ is not a Hall conductivity and  $\sigma_{xy}=\sigma_{yx}$ holds 
as in Eq. (\ref{eq:eq19})
in the case of zero magnetic field.\cite{Kubo} 
$\sigma_{xy}$ is finite when $\sigma_- \ne \sigma_+$. 
The sign of $\phi$ is chosen such that   $\phi < 0$ for $\sigma_{xy} > 0$ and        $\phi > 0$ for $\sigma_{xy} < 0$,  where 
   $0 <|\phi| < \pi/4$ for $\sigma_y > \sigma_x$ and  
   $\pi/4  <|\phi| < \pi/2$ for $\sigma_x > \sigma_y$. 

In terms of the conductivity , the resistivity is given by 
\begin{eqnarray}
\begin{pmatrix}
  \rho_x  & \rho_{xy}  \\
  \rho_{xy}  & \rho_y 
\end{pmatrix}  
=
\frac{1}{\sigma_x\sigma_y - \sigma_{xy}^2}
\begin{pmatrix}
  \sigma_y  & \sigma_{xy}  \\
  \sigma_{xy}  & \sigma_x 
\end{pmatrix}  
 \; .
 \label{eq:eq_rho}
\end{eqnarray}

\subsection{Spin susceptibility}

The magnetic (spin) susceptibility 
  by the Zeeman effect  is calculated as follows. 
Using the component of the wave function $d_{\alpha \gamma}$ 
 in Eq.~(\ref{eq:eq6c}), 
 the spin response function per spin
  is calculated as\cite{Katayama_EPJ} 
\begin{eqnarray}
\chi_{\alpha \beta} &=&  
 - \frac{1}{N}\sum_{k_x,k_y} \sum_{\gamma=1}^4 \sum_{\gamma'=1}^4   
  \frac{f(E_{\gamma}(\bk)) - f(E_{\gamma'}(\bk))} 
  {E_{\gamma}(\bk)-E_{\gamma'}(\bk)} \nonumber \\
& & \times  d_{\alpha \gamma}(\bk)^* d_{\beta \gamma}(\bk)
 d_{\beta \gamma'}(\bk)^* d_{\alpha \gamma '}(\bk) 
 \; .
  \label{eq:chi}
\end{eqnarray}
The local  magnetic susceptibility at the $\alpha$ 
 site,  $\chi_{\alpha}$,  is obtained as 
\begin{eqnarray}
\chi_{\alpha}(T) 
  &=&
    \sum_{\beta} \chi_{\alpha \beta} \nonumber \\ 
  &=& 
  -  \frac{1}{N}\sum_{\bk} \sum_{\gamma}
    \left(   \frac{\partial f (E_{\gamma}(\bk))}
                 {\partial E_{\gamma}(\bk)}  \right) 
   d_{\alpha \gamma}(\bk) d_{\alpha \gamma}(\bk)^* 
                                            \nonumber \\
&=&   - \int_{-\infty}^{\infty} {\rm d} \omega 
         \frac{\partial f(\omega)}{ \partial \omega} D(\omega) 
 d_{\alpha \gamma}(\bk) d_{\alpha \gamma}(\bk)^* 
\; . 
  \label{eq:chi_a}
\end{eqnarray} 
  The total magnetic susceptibility
  $\chi^{\rm total}(T)$, is obtained as 
\begin{eqnarray}
\chi^{\rm total}(T) =  \sum_{\alpha} \chi_{\alpha}(T) = 
    - \int_{-\infty}^{\infty} {\rm d} \omega 
         \frac{\partial f(\omega)}{ \partial \omega} D(\omega) \; ,  
  \label{eq:eq21}
\end{eqnarray}
 where  
 $\sum_{\alpha}d_{\alpha \gamma}(\bk) d_{\alpha \gamma}(\bk)^* = 1$
Note that 
the difference  between Eqs. (\ref{eq:chi_a}) and (\ref{eq:eq21}) 
is a factor  $d_{\alpha \gamma}(\bk) d_{\alpha \gamma}(\bk)^* $, which projects    $\chi^{\rm total}(T)$ into the respective molecular site.

\section{Results}
We calculate the conductivity for the TB model with transfer energies shown in Table~\ref{Transfer_alpha}. The direction of molecular stacking is given by the $y$ ($a$) axis, and that perpendicular to the stacking is given by the $x$ ($b$) axis (Fig.~\ref{fig1}), where the configurations of NN and NNN transfer energies are shown in Ref.~\citen{EPJB2020}.  
Table \ref{Transfer_alpha} shows the transfer energy $w_{h,s=\pm s'}$, which becomes complex in the presence of SOC, while it was treated as a real quantity for simplicity in the previous work.\cite{EPJB2020}
 The transfer energies were obtained from overlaps between maximally localized Wannier functions (MLWF) at each molecule,  generated using
 the   \texttt{wannier90} code~\cite{w90}. 
To construct MLWF, we selected eight bands close to the Fermi level (four bands made up of HOMO of BEDT-TTF molecule with up and down spins),  calculated using full-relativistic pseudopotentials with plane-wave basis sets, implemented in ~{\sc Quantum ESPRESSO}~\cite{QE}. The computational details are shown in Ref.~\cite{EPJB2020}.
 
\begin{figure}[tb]
\begin{center}
\includegraphics[width=6cm]{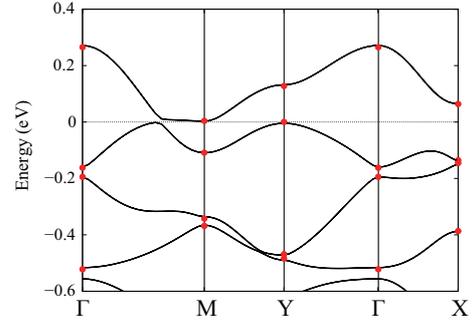}
\end{center}
\caption{(Color online)
Band structures including SOC effect along with the symmetric points in the first Brillouin zone, which denote time-reversal invariant momenta (TRIM) 
 given by 
 $\Gamma$ = (0,0,0),
 S~(M) =$ (\pi, \pi, 0)$,
 Y=$(0, -\pi,0)$,  and 
 X =$(\pi, 0, 0)$. 
The 2D vector is defined as 
$ \bk =  k_x(1,0) +k_y(0,1)$ 
 = ($k_x$, $k_y$).
The solid curves are obtained by the first-principles DFT method, 
 while the symbols are obtained from 
the TB model with Table \ref{Transfer_alpha}.
~\cite{EPJB2020}  
The symbols  agree with those from the DFT calculations (solid curves) within an energy scale of 0.01 eV.  
The energy zero is set to be the top of the valence bands~[$\tilde{E}^{\rm SO}_{3}$($\bk$) and $\tilde{E}^{\rm SO}_{4}$($\bk$)].
}
\setlength\abovecaptionskip{0pt}
\label{Fig_band_TRIM}
\end{figure}

In Fig.~\ref{Fig_band_TRIM},
 the energy spectrum is  shown, 
suggesting a good agreement between 
 the DFT calculation (solid curves) and the TB calculation 
 (symbols at TRIM).  
As shown later,  the imaginary part gives a significant contribution to 
  the conductivity, although the real part is enough for the DOS.   
  In the following calculations,  the conductivity is normalized by 
  $e^2/\hbar$.
\begin{table}[b]
\centering
\caption{Effective transfer energies and 
 site-dependent potential energies
in eV for $\alpha$-(BETS$_2$)I$_2$. 
$\Delta V_{B}$ and $\Delta V_{C}$ are the difference of site-potential energies of B and C molecular sites relative to the $A$ (and $A^{\prime}$) sites, respectively.  The definitions for transfer energies  are shown in Appendix A,
where $t_{\alpha s,\beta,s'} \rightarrow w_{h,s=\pm s'}$ and 
 $h = a1, \cdots, s4.$~\cite{EPJB2020},
}
\begin{minipage}{0.5\hsize}
\centering
\label{Transfer_alpha}
\begin{tabular}{crrrrrrr}
\hline\hline
$w_h$, $s =s'$ & Re($w_h$) & Im($w_h)$ \\ 
\hline
$a1$  & 0.0053  & 0.001302  \\
$a2$  & -0.0201  &  0 \\ 
$a3$  & 0.0463  & 0 \\ 
\hline        
$b1$  & 0.1389  & 0.00674 \\ 
$b2$  & 0.1583  & 0.007319 \\ 
$b3$  & 0.0649  & 0002037 \\ 
$b4$  & 0.0190  & -0.001202 \\ 
\hline        
$a1'$ & 0.0135  & 0 \\
$a3'$ & 0.0042  & 0 \\ 
$a4'$ & 0.0217  & 0 \\ 
\hline        
\hline        
$c1$  & -0.0024  & -0.000564 \\ 
$c2$  &  0.0063  & -0.000104 \\ 
$c3$  & -0.0036  & -0.00040 \\ 
$c4$  & 0.0013   &  0.00027 \\ 
\hline        
$d0$  & -0.0009  & 0 \\ 
$d1$  & 0.0104   & 0 \\ 
$d2$  & 0.0042   & -0000098 \\ 
$d3$  & 0.0059   & -0.000039 \\ 
\hline        
$s1$  & -0.0016  & -0.000172 \\ 
$s3$  & -0.0014  &  0 \\ 
$s4$  &  0.0023  &  0 \\
\hline
\hline
$\Delta$$V_B^{\rm{}}$ & -0.0047 & \\ 
$\Delta$$V_C^{\rm{}}$  &  - 0.0092  &\\   
\hline\hline
\\
\end{tabular}
\end{minipage}
\begin{minipage}{0.45\hsize}
\centering
\begin{tabular}{cr}
\hline
\hline
$w_h$,$s=-s'$\\ 
\hline
$b1_{so1}$  & -0.0020 \\ 
$b1_{so2}$  & 0.0020 \\
$b2_{so1}$  & -0.0019 \\
$b2_{so2}$  & 0.0019 \\ 
$b4_{so1}$  & -0.0008 \\ 
$b4_{so2}$  & 0.0008 \\ 
$c1_{so1}$  & 0.0007 \\ 
$c1_{so2}$  & -0.0007 \\ 
$c2_{so1}$  & 0.0003 \\ 
$c2_{so2}$  & -0.0003 \\ 
$c3_{so1}$  & 0.0006 \\
$c3_{so2}$  & -0.0006 \\
$c4_{so1}$  & 0.0001 \\ 
$c4_{so2}$  & -0.0001 \\  
\hline
\hline
\end{tabular}
\end{minipage}
\end{table}

\begin{figure}
  \centering
\includegraphics[width=3.5cm]{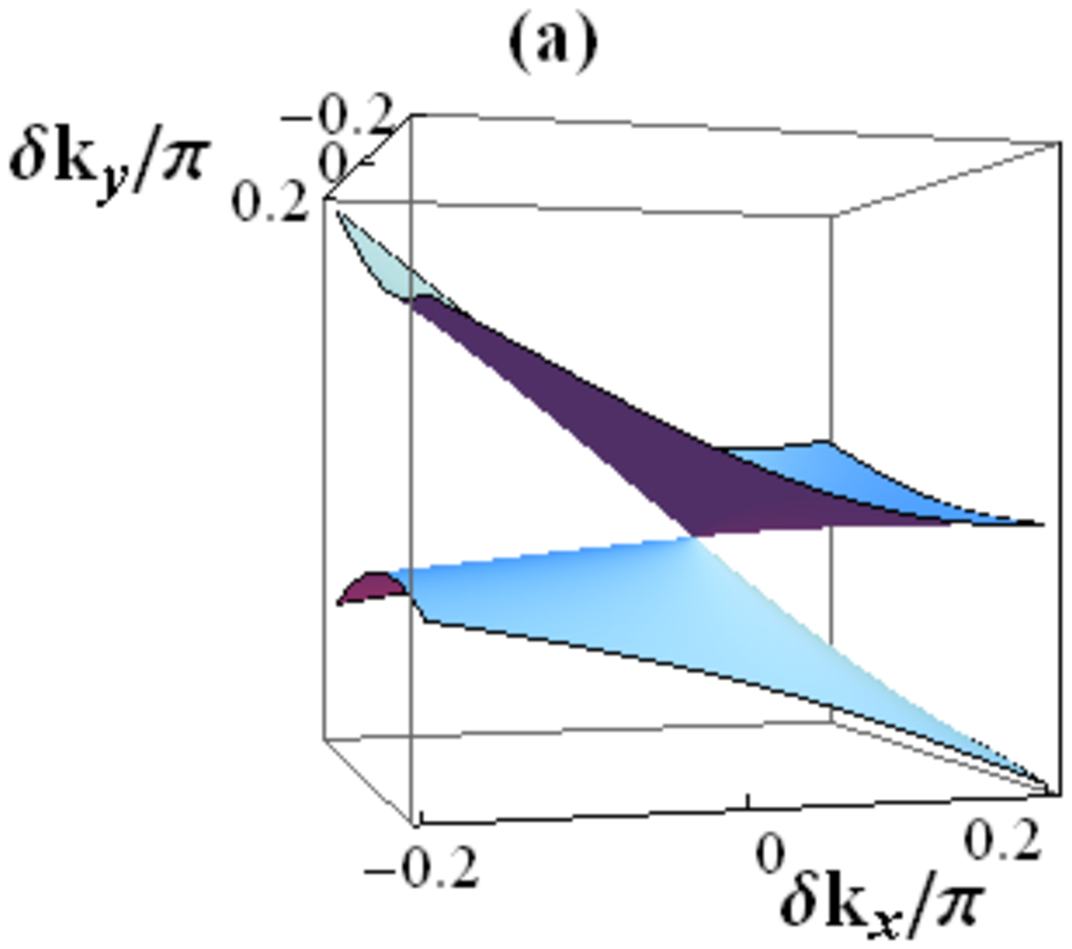}
\includegraphics[width=3.5cm]{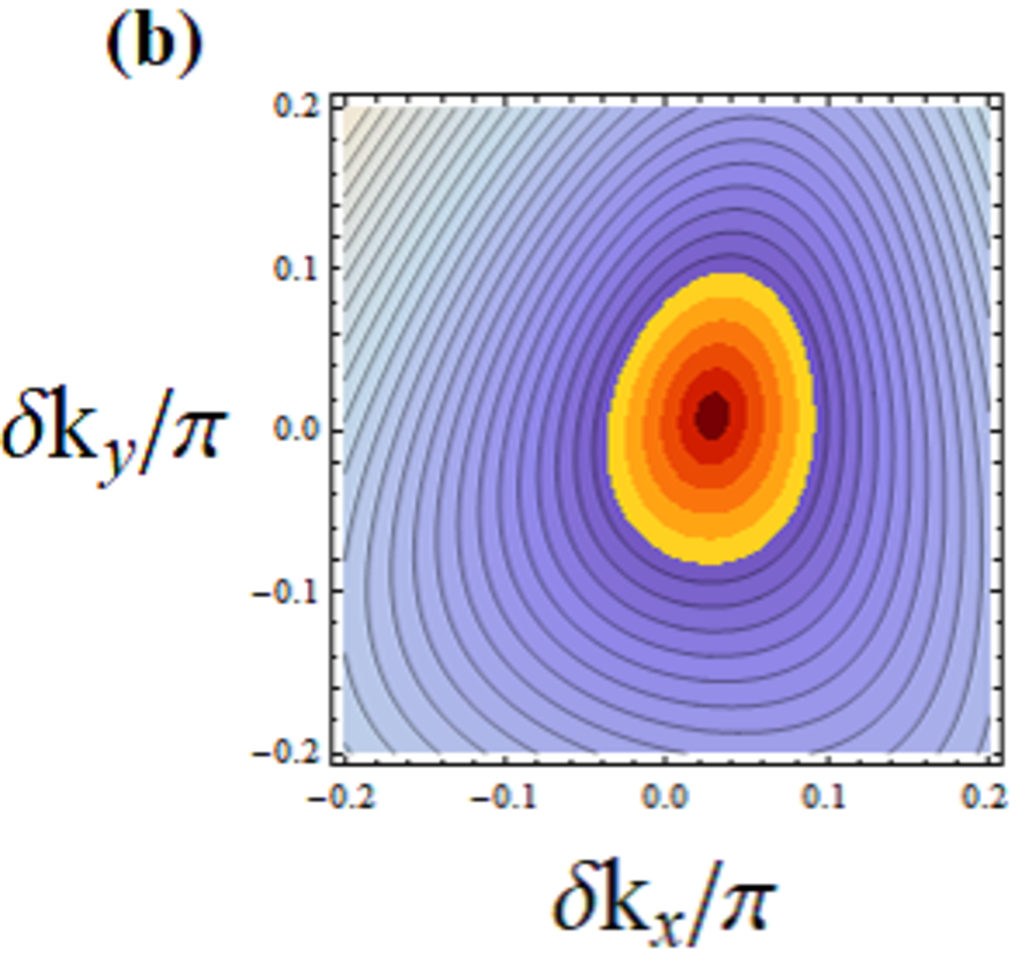}
 \\
\includegraphics[width=3.5cm]{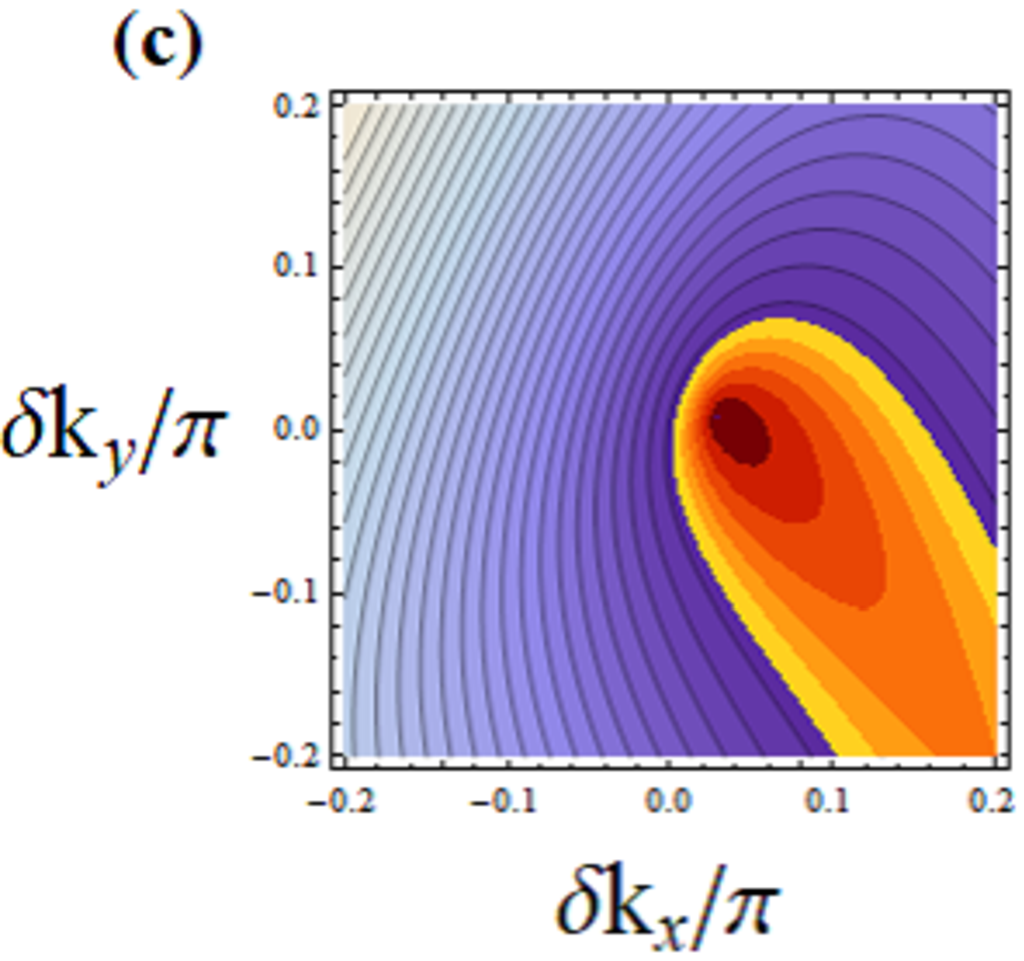}
\includegraphics[width=3.5cm]{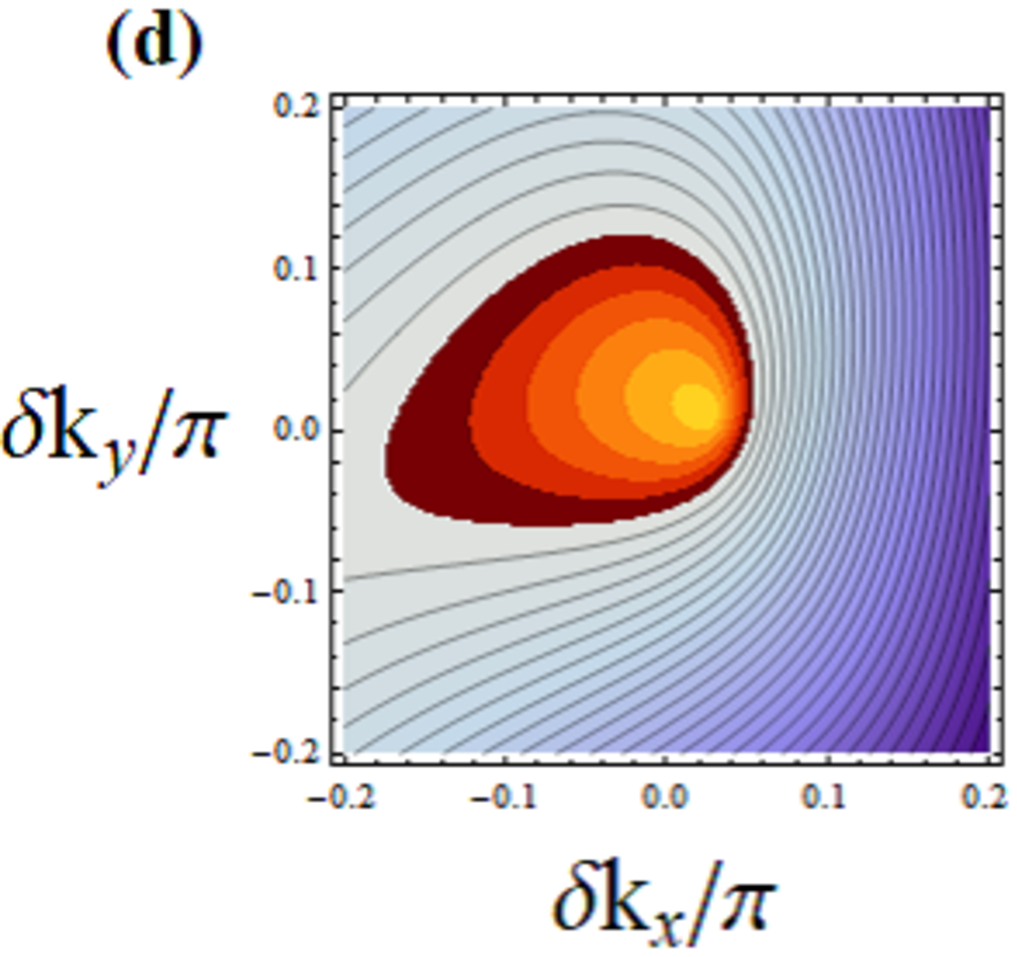}
\includegraphics[width=3.5cm]{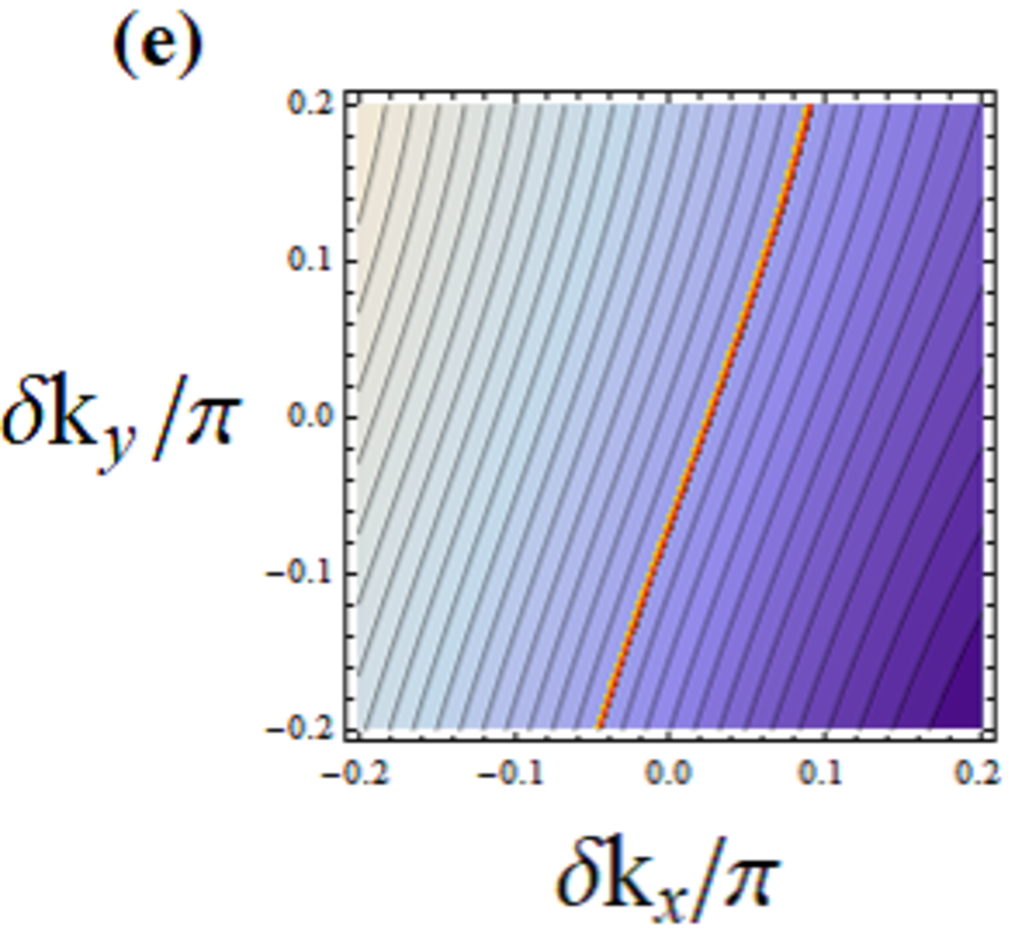}
 \\
     \caption{(Color online)
(a) Conduction and valence bands given by  
$E_1(\bk)$ (upper band) and $E_2(\bk)$ (lower band).
 Two bands contact at the Dirac points $\pm \bkD = \pm (0.72, -0.58)\pi$
  with an energy $\eD = \mu = 0.1684$. 
 Here we use $\delta \bk = \bk - \bkD = (\delta k_x, \delta k_y)$. 
 (b) Contour plots of   $E_1(\bk) - E_2(\bk) (< 0.144)$.
 The outermost bright line (OBL) corresponds to 
$E_1(\bk) - E_2(\bk)$ =  0.03. 
 (c) Contour plots  of $E_1(\bk) - \eD$ with the range [0, 0.19], where  
   the Dirac point exists at (0,0). 
 The OBL corresponds  to 
  $ E_1(\bk) - \eD$  = 0.01.  
 (d) Contour plots of $E_2(\bk) - \eD$ with the range [-0.074, 0]. 
  The Dirac point exists at (0,0). The OBL  corresponds to $ E_2(\bk) - \eD $ = -0.01. 
 (e) Contour plots of $E_1(\bk) + E_2(\bk) - 2\eD$ with the range  [-0.0636, 0.0901].     The bright line denotes  $E_1(\bk) + E_2(\bk) - 2\eD = 0$.
  }
\label{fig3}
\end{figure}

 Figure \ref{fig3}(a) shows two bands of $E_1(\bk)$ and $E_2(\bk)$ 
 as the function of $\delta \bk = \bk -\bkD$, where 
  Dirac points are given by $\pm \bkD = \pm (0.72, -0.58)\pi$ 
 with an energy $\eD$ = $\mu = 0.1684$ 
 corresponding to the three-quarter-filled band.
The ranges of the energy of the conduction and valence 
  bands $E_1(\bk)$ and $E_2(\bk)$  are  given by 
   $0 <E_1(\bk)-\eD < 0.17$ and $-0.074 < E_2(\bk)-\eD< 0$, respectively. 
Such  ZGS shows  the relation 
  $E_2(Y) < \eD <  E_1(M)$, where  $\Gamma$, X, Y, and M are TRIMs  given by  
 $\Gamma=(0,0)\pi$, $X=(1,0)\pi$,
 $Y=(0,1)\pi$, and $ M=(1,1)\pi$, respectively.
Figure \ref{fig3}(b) shows  
  contour plots of $ E_1(\bk)-E_2(\bk)$  around  $\bkD$. 
 The contour lines form anisotropic circles, 
suggesting that 
 the velocity of the  Dirac cone is  large for $k_x$ direction. 
In fact, 
 $ E_1(\delta \bk)-E_2(\delta \bk) = 2 \bm{v}\cdot \delta \bk$ for small
 $|\delta \bk|$ gives $v_x = 0.075$ and  $v_y=0.053$, which 
 are compared with those of 
 $\alpha$-(ET)$_2$I$_3$ ( $v_x = 0.053$ and  $v_y=0.043$).
\cite{Katayama_EPJ,Suzumura2021_JPSJ}   
Figure \ref{fig3}(c) 
 shows $E_1(\bk) - \eD$. 
The Dirac point is located at 
 $(\delta k_x, \delta k_y)$ = (0,0).
 This contour  
suggests a tilted  Dirac cone and  
 shows a slight  deviation from the ellipse.
 In terms of a tilting velocity $v_t$ 
   and the corresponding velocity of the Dirac cone $V$, 
  the tilting parameter is estimated as $\eta = v_t/V \simeq 0.8$, 
 which is nearly the same as that of 
$\alpha$-(ET)$_2$I$_3$.\cite{Kobayashi2008} 
It is also found that 
 the cone shows  
 a slight rotation   clockwise from the  $k_x$ axis, 
 in contrast to that of $\alpha$-(ET)$_2$I$_3$ under hydrostatic pressure.
\cite{Suzumura2021_JPSJ}
Figure \ref{fig3}(d) 
 shows $E_2(\bk) - \eD$. 
The Dirac point is located at (0,0).
 The contour of $E_2(\bk) - \eD$  
 also shows  a tilted  Dirac cone and a slight deviation from the ellipse. 
We define  a phase  $\phi_1 (<0)$ (
$\phi_2 $) as 
  a tilting angle of $E_1(\bk)$ ($E_2(\bk)$)  measured from the $k_x$  axis. 
Since $E_1(\bk)$ and $E_2(\bk)$ form a pair of Dirac cones, 
 $\phi_2 - \phi_1 = \pi$ for $\bk$ in the limit of the  Dirac point. 
 The deviation from the limiting value 
 increases   with increasing $|\delta \bk|$. 
Figure \ref{fig3}(e)  shows a bright color line 
  of $E_2(\bk) + E_1(\bk) = 2 \eD$, on  which 
  the Dirac point is located, i.e., the apex of the Dirac cone. 
 Thus, the line is almost perpendicular to the tilting axis, which rotates clockwise from the $k_x$ axis as discussed later. 

\begin{figure}
  \centering
\includegraphics[width=7cm]{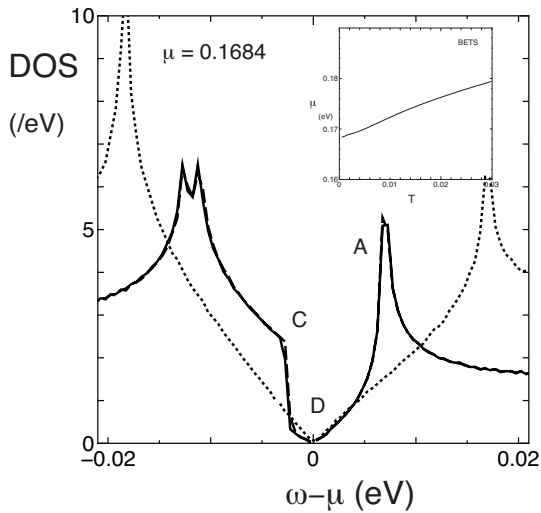}
     \caption{
DOS as a function 
 of $\omega -  \mu=\omega$. 
The inset denotes chemical potential ($\mu$)  
as a function of temperature ($T$). 
The solid (dot-dashed) line represents DOS 
  with (without) the imaginary part of the transfer energy.
The dotted line corresponds to DOS for $\alpha$-(ET)$_2$I$_3$ 
under hydro-static  pressure.~\cite{Katayama_EPJ} 
}
\label{fig4}
\end{figure}

Figure \ref{fig4} shows DOS as a function of $\omega - \mu$, 
 where the inset denotes the $T$  dependence 
 of the chemical potential $\mu(T)$.
 $\mu$ is the chemical potential at $T$=0. 
The  van Hove singularities exist 
at  C [$E_2(Y)$],  peaks below C
 ( an  intermediate region between  $\bkD$ and Y), and 
  A [$E_1(M)$]. 
 With increasing $T$,   $\mu$ varies almost linearly. 
 The increase in $\mu$  occurs since 
 the van Hove singularity  below the chemical potential  
has a large peak  compared with the above one. 
 The DOS close to the chemical potential  shows 
 a linear dependence for $\omega - \mu$. 
 However,  the range is narrow compared with 
that expected from  Figs.~\ref{fig3}(a) and \ref{fig3}(b).
 Such a difference  is ascribed to 
 the effect of  A ($E_1$(M)) for $\omega - \mu > 0$ and 
 the effect of C ($E_2$(Y)) for $\omega - \mu < 0$. 
The energy at the respective TRIM is close to that at the Dirac point.
Moreover, the former(latter)  shows a singularity due to a saddle point 
(a maximum). The behavior at the C point is in contrast  with 
 that of  $\alpha$-(ET)$_2$I$_3$, which exhibits a saddle point due to 
   $E_2$(Y) being much lower than  $E_2(\eD)$.

\begin{figure}
  \centering
\includegraphics[width=6.5cm]{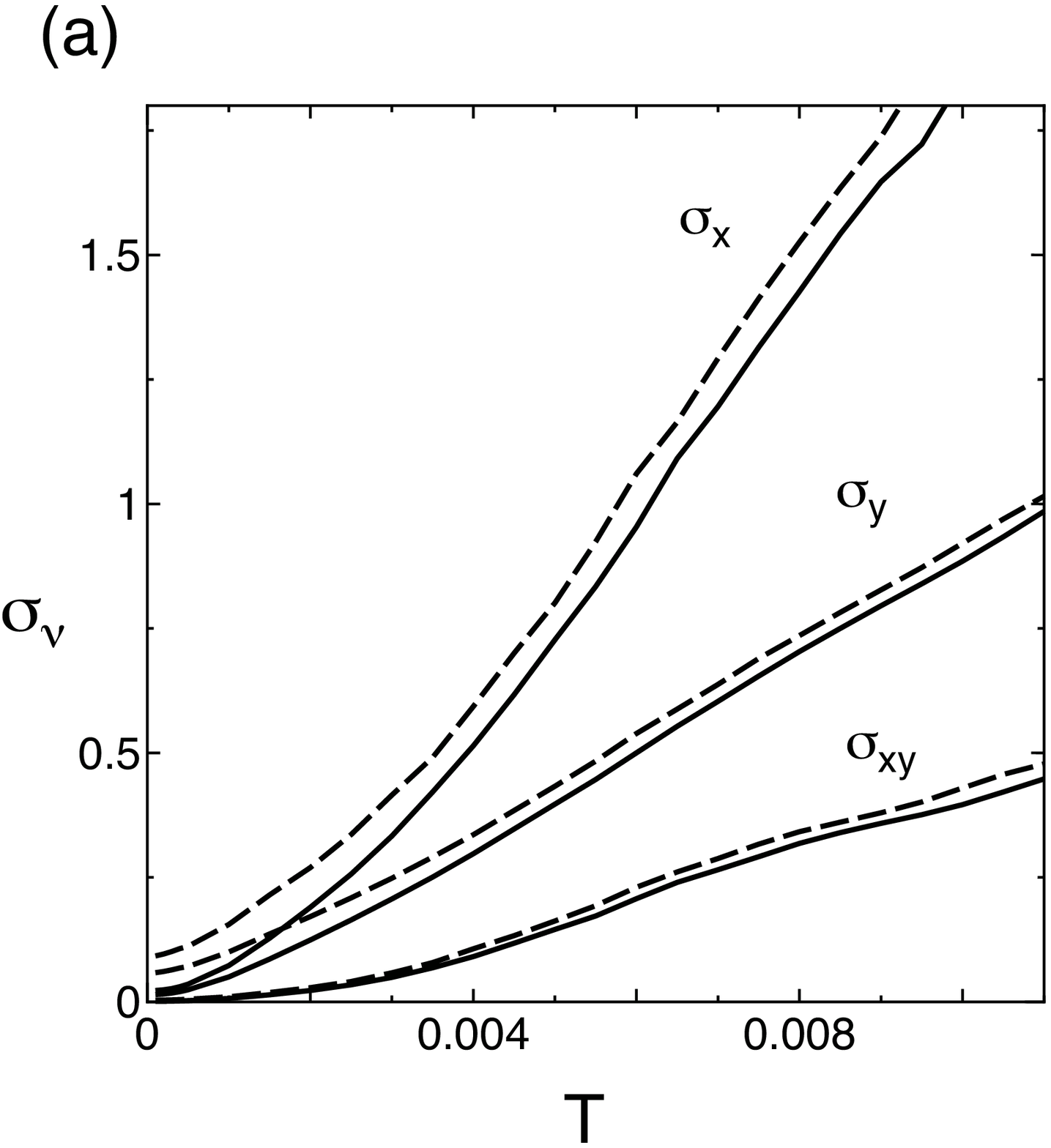}
\includegraphics[width=6.5cm]{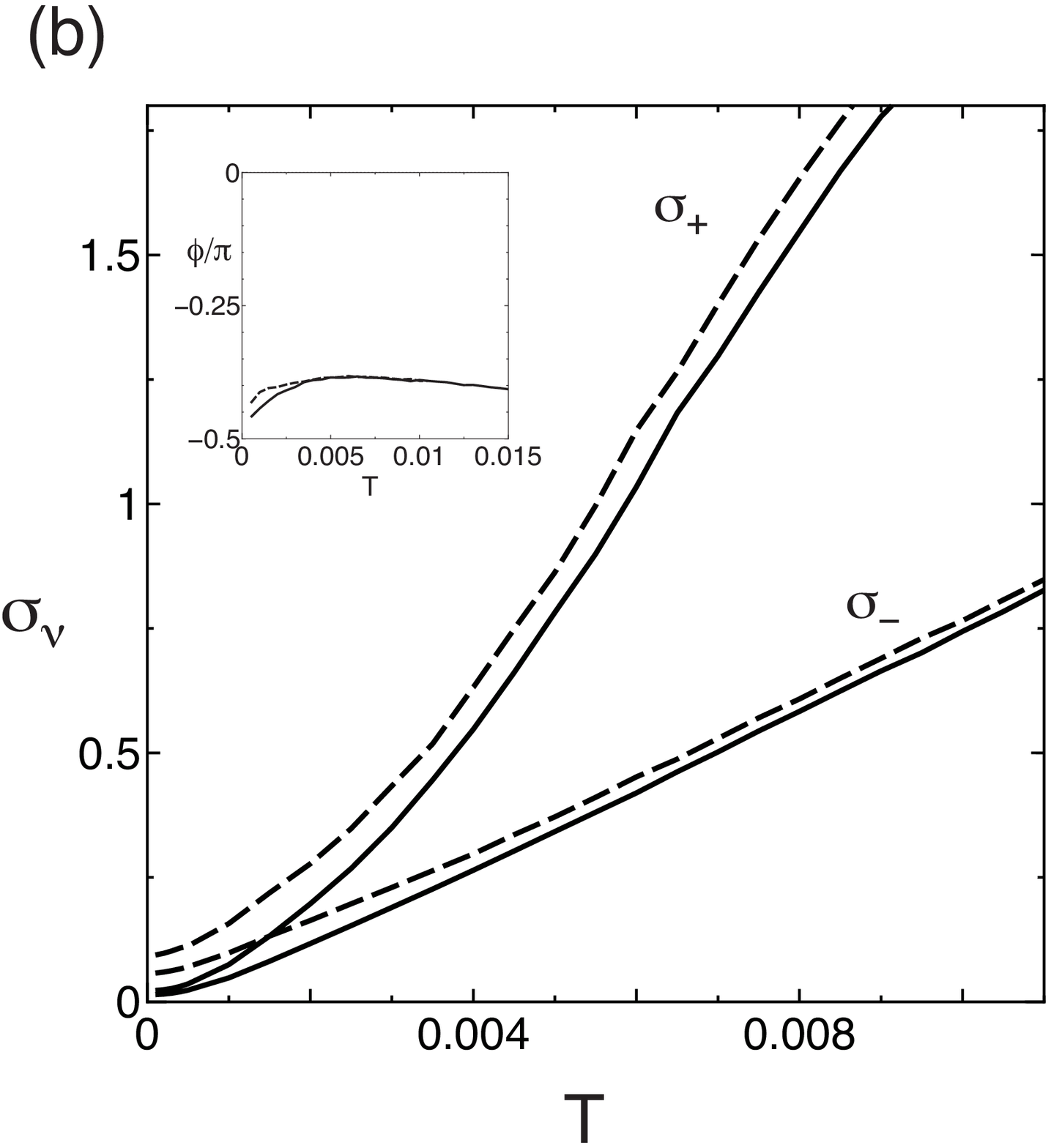}
\includegraphics[width=6.5cm]{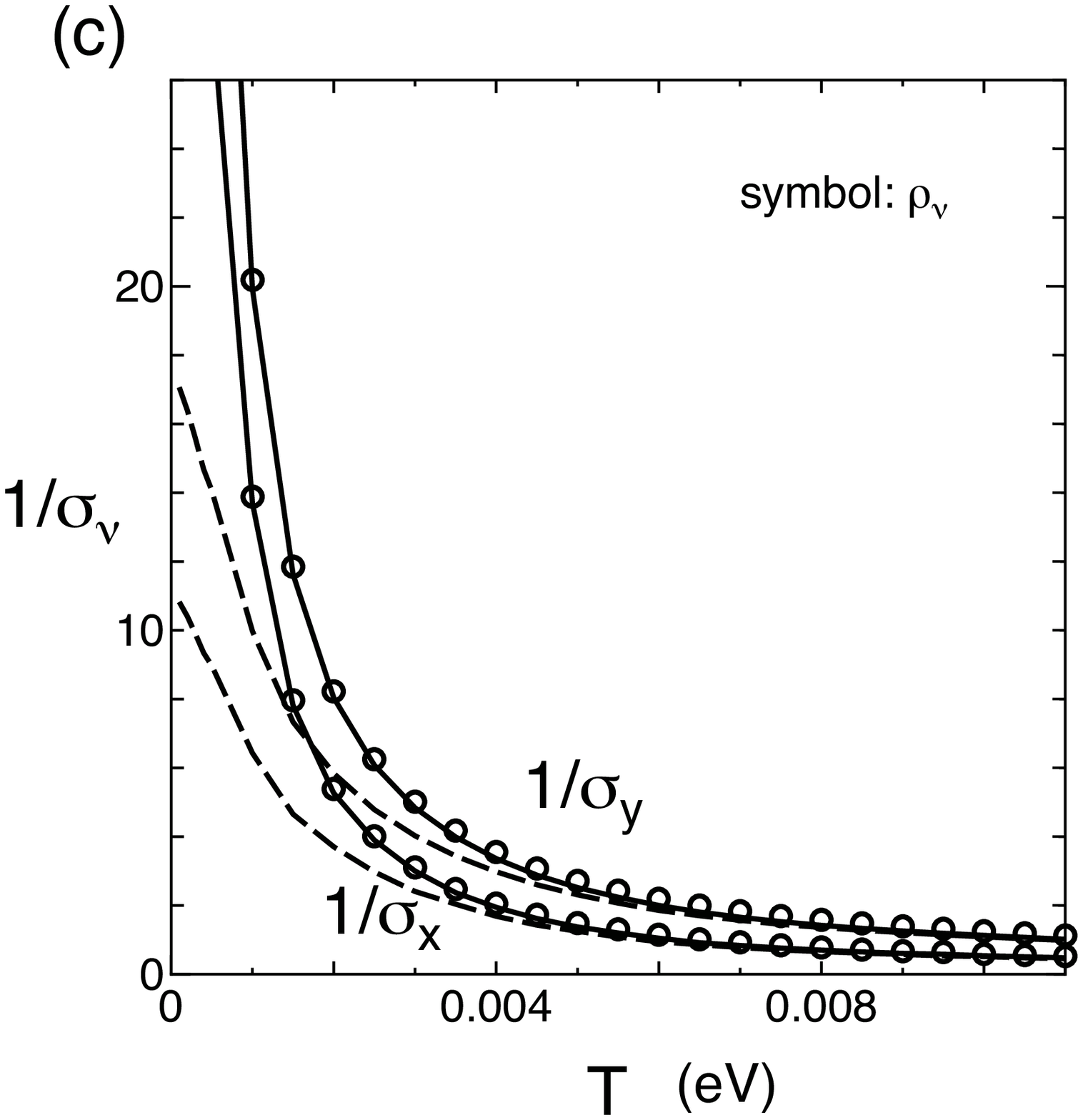}
     \caption{
$T$ dependence of conductivity and  resistivity 
in the absence of  the e--p interaction 
with fixed $\Gamma$ = 0.0005.
 The dashed (solid) line corresponds to  
 the case without (with) the imaginary part due to  SOC.
Figure \ref{fig5}(a) shows 
$\sigma_x$, $\sigma_y$, and $\sigma_{xy}$.
Figure \ref{fig5}(b) shows $\sigma_\pm$.  
 Principal values of $\sigma_{-}$  and $\sigma_{+}$
 are given by Eqs.~(\ref{eq:21b}) and (\ref{eq:eq21c}), respectively, 
 while $\phi$ is given by Eq.~(\ref{eq:eq21a}).
 The inset shows the phase $\phi$, which is  an angle of 
 the principal axis of $\sigma_-$  measured from the $k_x$ axis. 
Figure \ref{fig5}(c) shows the resistivity given by 
 $1/\sigma_{\nu}$, where  the 
 symbol shows $\rho_{\nu}$ obtained from Eq.~(\ref{eq:eq_rho}). 
}
\label{fig5}
\end{figure}

Now we examine the conductivity and resistivity 
 using Eqs.~(\ref{eq:sigma}) and (\ref{eq:eq19})
  with $\Gamma$ = 0.0005.
We only calculate  for the case of the SOC only with the same spin, i.e.,  
 by discarding $w_{s=-s'}$, which results in the insulating gap 
 of $\simeq$  1 meV. Thus the present result is applied for $0.001 < T$, 
 where the resistivity enhancement at low temperature 
 is  still expected. 
First, we examine the case without the e--p interaction.
Figures \ref{fig5}(a), \ref{fig5}(b) and \ref{fig5}(c)   show 
the temperature dependence of 
 conductivity of  $\sigma _\nu$ ($\nu = x, y,xy, +, -$), 
$\sigma _\nu$ ($\nu = +, -$) and 
  $1/\sigma _\nu$ ( =   $\rho_\nu$), respectively.
The case of transfer energy of only real part (dashed line)
  is compared with that of both real and imaginary parts 
 (solid line). 

In Fig.~\ref{fig5}(a), one finds a relation $\sigma_x > \sigma_y$  
   for arbitrary $T$. 
This comes from the fact that the effect of  
 the anisotropy of velocity, $v_x/v_y \simeq 1.4$ gives a significant  effect 
 compared  with   the tilting, $\eta \simeq 0.8$. 
  This can be understood from a fact that 
 the ratio of $\sigma_x/\sigma_y$ in the limit of the Dirac cone  
 is  proportional to a product of   $(v_x/v_y)^2$ and  
    $\sqrt{1-\eta^2}\eta^{-1} \sin^{-1}\eta$.
\cite{Suzumura_JPSJ_2014}    
 Since $\sigma_{\nu}$ obtained for transfer energy in the presence of  
 the imaginary part (solid line)
 is  smaller than that with only real part (dashed line), 
 it turns out that 
  the SOC reduces the conductivity. 
In the presence of SOC ( solid curve), 
 the difference between $\sigma_x$ and $\sigma_y$ at low temperature 
 becomes negligibly small while the case without SOC (dashed line) 
 shows a clear difference  even at low 
 temperature as seen also in the case of $\alpha$-(ET)$_2$I$_3$.\cite{Suzumura2021_JPSJ}
  
In Fig.~\ref{fig5}(b), principal values $\sigma_{\pm}$  
are shown. 
The inset denotes  the rotation angle $\phi$ of  $\sigma_+$ measured from 
the  $y$ axis.
Note that  
 the axis for $\sigma_+$ is  perpendicular to 
  the axis of the cone  when  the  velocity of the cone 
is isotropic.~\cite{Suzumura_JPSJ_2014}
However   $\phi/\pi (< -0.25) $ in the inset  shows that  
  there is a large deviation of the axis  of $\sigma_+$ 
 from that expected by the tilting (see Fig.~\ref{fig3}(c)). 
This suggests $\sigma_+$, which  is mainly determined 
 by the anisotropy of the velocity, i.e., $v_x > v_y$. 
The axis for $\sigma_+$   rotates clockwise 
 in accordance with   $\sigma_{xy}>0$. 
It is found that  $\sigma_+ \simeq \sigma_x$  and  
  $\sigma_- \simeq \sigma_y$ due to small $\sigma_{xy}$. 
Thus the magnitude of the dominant conductivity is given by $\sigma_x$, 
 and the direction is relatively close to the $x$-direction.
We note that $\sigma_{+}$ at low temperatures is convex downward, 
which contrasts  that of  $\alpha$-(ET)$_2$I$_3$.
This is understood from 
 the comparison between the  solid line [$\alpha$-(BETS)$_2$I$_3$] 
and dotted line [$\alpha$-(ET)$_2$I$_3$] 
  in Fig.~\ref{fig4}, where 
 the region of the linear  dependence for DOS of  $\alpha$-(BETS)$_2$I$_3$ 
 is narrower 
 than that of $\alpha$-(ET)$_2$I$_3$.

Figure \ref{fig5}(c) shows  $T$ dependence  of 
 $1/\sigma_{\nu}$ ($\nu = x$ and $y$), where the dashed line 
 corresponds to  the transfer energy with 
 an only real part. 
It increases gradually with decreasing temperature.
The solid line corresponding to  $1/\sigma_{\nu}$ with  both real and imaginary parts shows a noticeable  enhancement at low temperatures. 
 The symbols denote  $\rho_{\nu}$ ($\nu = x$ and $y$) 
 obtained from Eq.~(\ref{eq:eq_rho}).
The difference between  $1/\sigma_{\nu}$ and  $\rho_{\nu}$ 
 is negligibly small 
 due to  the  small $\sigma_{xy}$.

\begin{figure}
  \centering
\includegraphics[width=6.5cm]{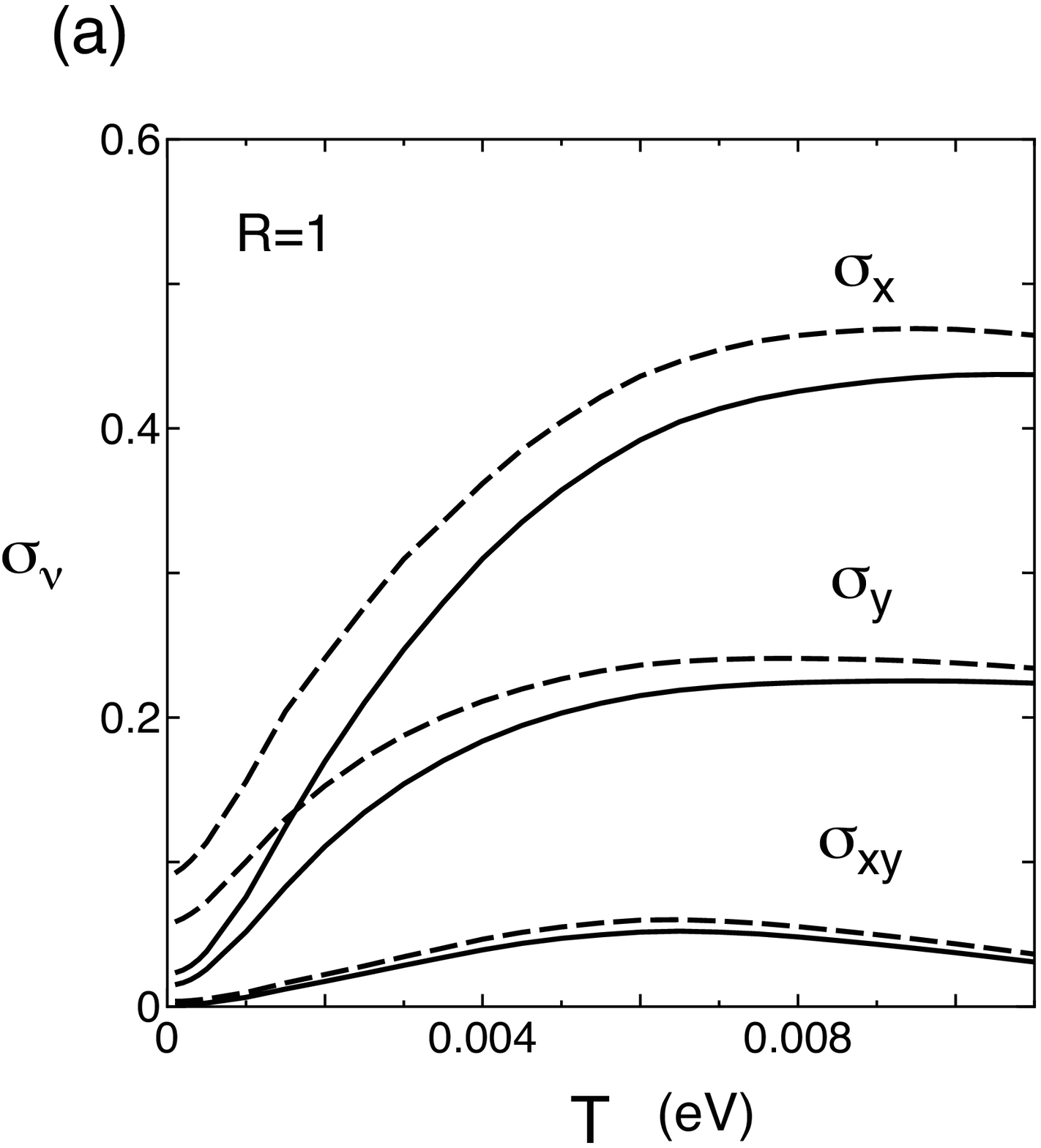} 
\includegraphics[width=6.5cm]{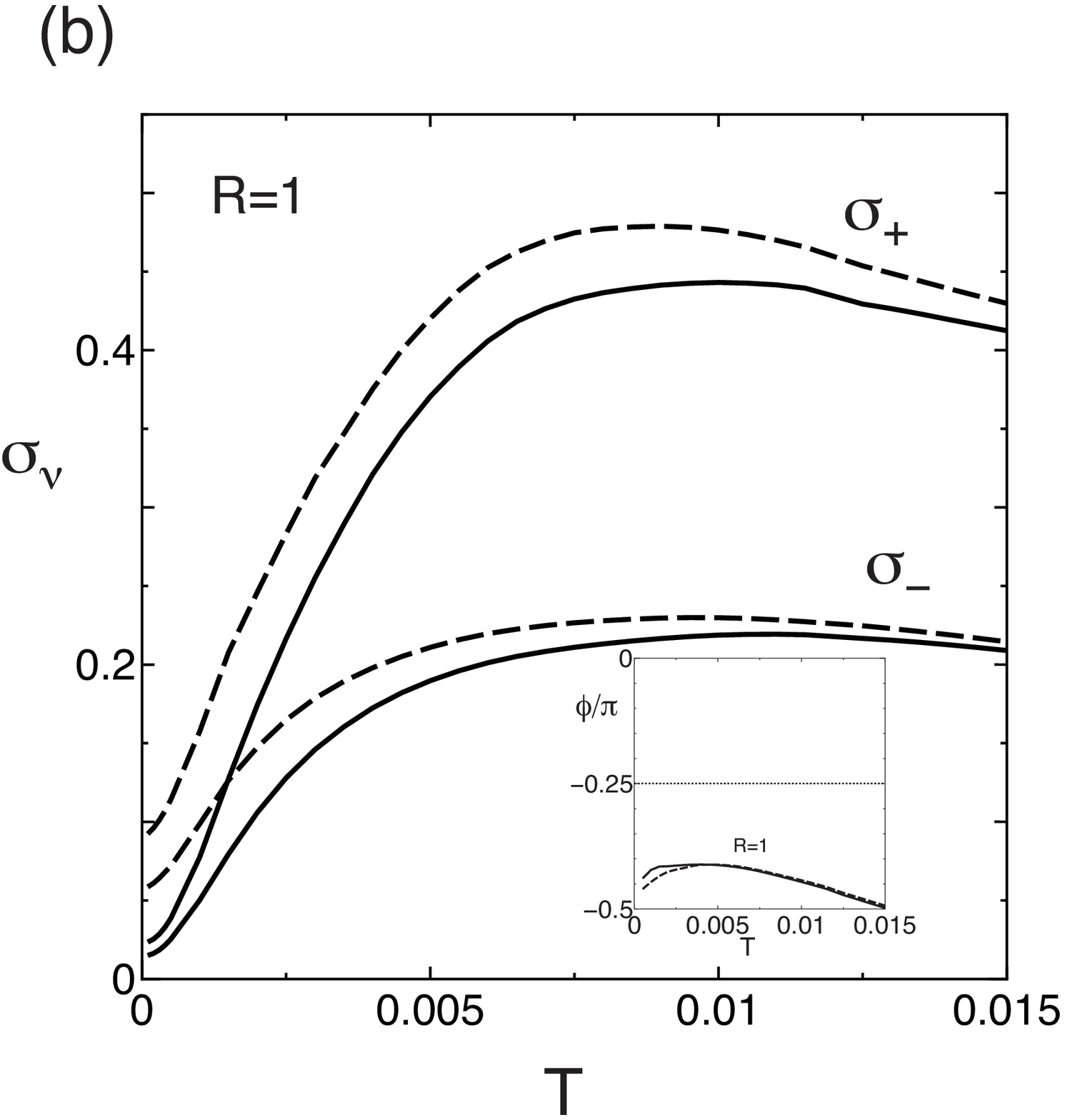} 
\includegraphics[width=6.5cm]{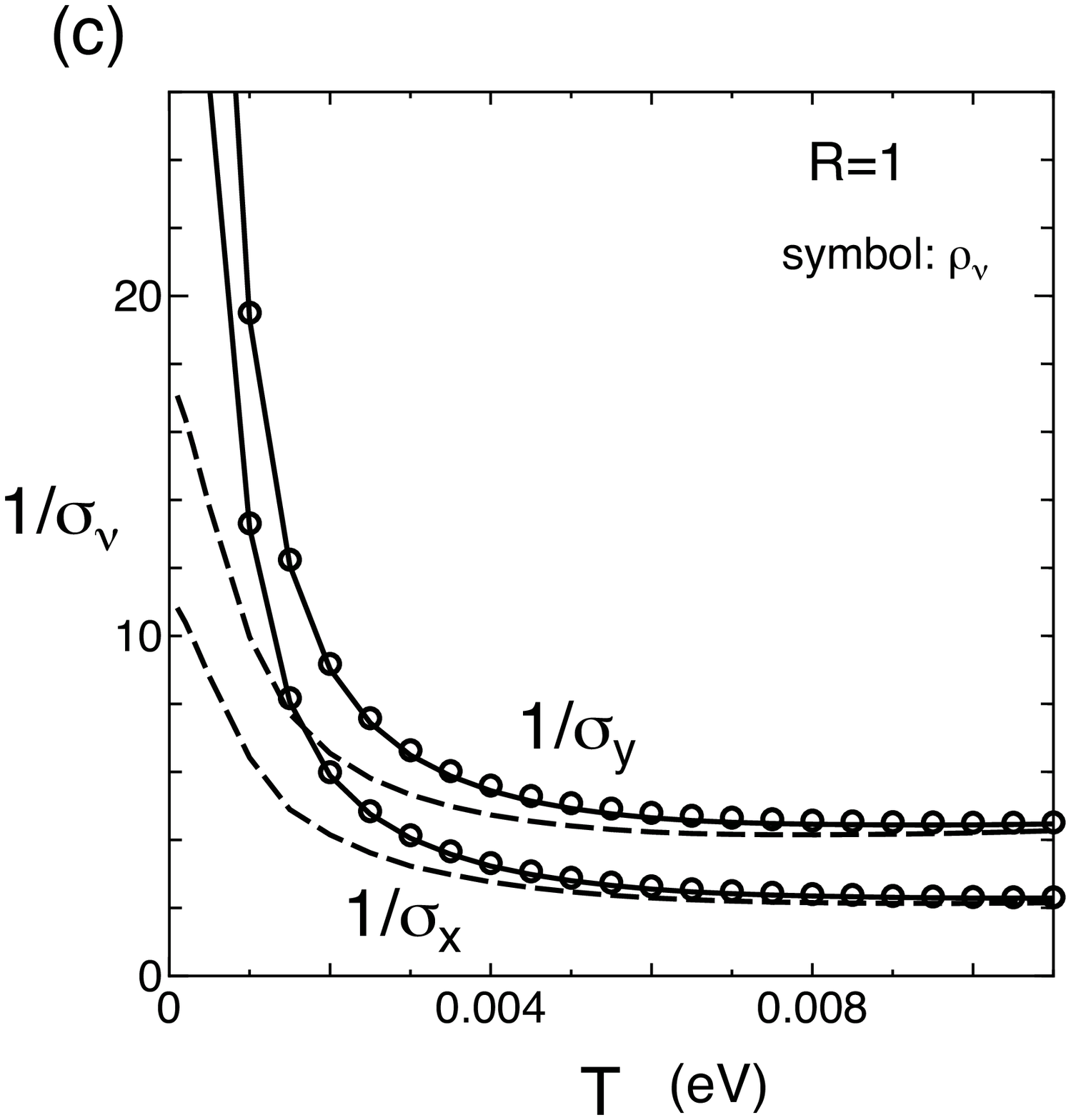} 
     \caption{
$T$ dependence of conductivity  and 
 resistivity in the presence of  the e--p interaction 
with fixed $\Gamma$ = 0.0005. 
 The normalized   e--p coupling constant 
is taken as  $R$=1, 
   which is defined by Eq.~(\ref{eq:eq16b}).
 The dashed (solid) line corresponds to  
 the case without the SOC  (with the SOC).
Figure \ref{fig6}(a) shows 
$\sigma_x$, $\sigma_y$, and $\sigma_{xy}$.
Figure \ref{fig6}(b) shows $\sigma_\pm$.  
 The inset shows the phase $\phi$, which is  an angle of 
 the principal axis of $\sigma_-$  measured from the $k_x$ axis. 
Figure \ref{fig6}(c) shows the resistivity given by 
 $1/\sigma_{\nu}$, where  the 
 symbol shows $\rho_{\nu}$  obtained from Eq.~(\ref{eq:eq_rho}). 
}
\label{fig6}
\end{figure}

\begin{figure}
  \centering
\includegraphics[width=7cm]{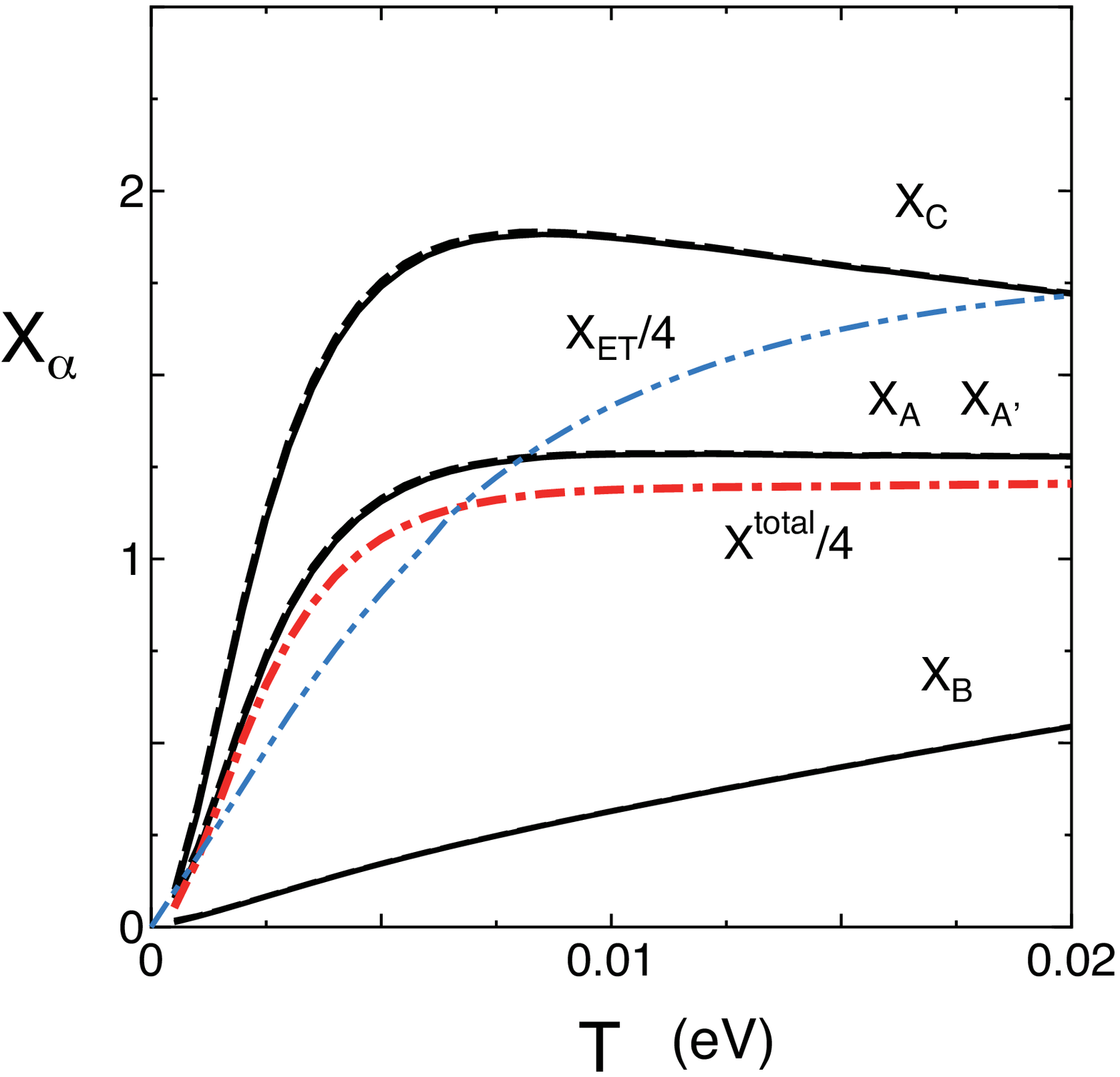} 
     \caption{(Color online)
Local magnetic susceptibility for A(=A'), B, and C molecular sites.
The dashed line (solid line) is obtained  
 for the transfer energy with  real (complex).
 The dot-dashed line denotes $\chi^{\rm total}/4$ 
 given by Eq.~(\ref{eq:eq21}).
The 2 dot-dashed line ($\chi_{\rm ET}$) corresponds to $\chi^{\rm total}$ 
 for $\alpha$-(ET)$_2$I$_3$,~\cite{Katayama_EPJ}
 which is compared with that for BETS (dot-dashed line).
}
\label{fig7}
\end{figure}

 As shown in Fig.~\ref{fig5}(a),
  $\sigma_{\nu}$ increases monotonically  as a function of $T$, 
 while the conductivity   shows 
   nearly constant  at high temperatures.
~\cite{Inokuchi1995_BCSJ68}
Such an exotic $T$ dependence of $\sigma_{\nu}$ is examined   
  by taking account of  the e--p interaction, 
      which is expected to reduce $\sigma_{\nu}$. 
By using  Eq.~(\ref{eq:damping}) and (\ref{eq:eq16a}),
   we calculate $\sigma_{\nu}$  of Eq.~(\ref{eq:sigma}), 
     where  $\Gamma$ in the absence of the e--p interaction 
      is replaced by 
        $\Gamma_{\g} (= \Gamma + \Gamma_{\rm ph}^{\g})$. 
Owing to the $T$ dependence of   $\Gamma_{\rm ph}^{\g}$, 
   $\Gamma$ is dominant at low $T$,  
     whereas $\Gamma_{\rm ph}^{\g}$ is dominant at high $T$.
Note that  such  crossover with increasing $T$ depends on $R$.

Figure \ref{fig6}(a) shows the  $T$ dependence of 
    $\sigma_{\nu}$ ($\nu = x, y,$ and $xy$) 
       in the presence of the e--p interaction 
  with a choice of $R = 1$. 
 The effect of the e--p interaction appears when   
 $\sigma_{\nu}$ deviates from the $T$-linear behavior. 
  Compared with  $\sigma_{\nu}$ with $R$ = 0  (Fig.~\ref{fig5}(a)),
 $\sigma_{\nu}$  is reduced  noticeably. 
  At temperatures above $T \sim 0.008$,  $\sigma_x$ 
    becomes nearly constant,  
      while  $\sigma_y$  shows such behavior at lower temperatures. 
Such a  constant behavior of the conductivity 
 is understood as follows.
With increasing $T$,   
$\sigma_{\nu}$ without the e--p interaction ($R = 0$) 
  increases linearly owing  to 
 the DOS obtained from the Dirac cone.
 For $R \not= 0$, the linear increase is suppressed  
 at finite temperatures,  since   
 the effect of the acoustic phonon increases with increasing  temperatures.
The electron is scattered by both normal impurity ($\Gamma$) 
 and the e--p interaction ($\Gamma_{\rm ph}^{\gamma}$), and 
  the  latter becomes dominant  at high temperatures 
    as seen from  Eq.~(\ref{eq:eq16a}).
 However, for  the case of the  Dirac cone close to the  three-quarter-fil
led  band,
  the effect of the  e--p scattering is  strongly  reduced 
 owing  to a constraint
  by the energy-momentum conservation.\cite{Suzumura_PRB_2018} 
   Thus  a nearly constant behavior or  a broad maximum in $\sigma_{\nu}$,
      is obtained 
owing  to 
  a competition between the enhancement by  DOS  and     
 the suppression by the e--p interaction in the   Dirac electron system.  

Here we mention the condition for   $\Gamma_{\rm ph }$ 
 of Eq.~(\ref{eq:eq16a}),~\cite{Suzumura_PRB_2018} which has been estimated 
  for the acoustic phonon with an energy $v_{s}q$.
Since the velocity $v_s$ of  the acoustic phonon is   much smaller 
 than $v$  of Dirac cone, 
  the energy-momentum conservation  allows the classical treatment 
 for  a phonon distribution function 
   due to $v_s q \ll vq \simeq T$. 
Furthermore, the numerical estimation 
  shows that  
Eq.~(\ref{eq:eq16a}) is proportional to the energy $\xi_{\bk,\gamma,s}$
 for  $|\xi_{\bk,\gamma,s}|/\Gamma_0 < 20$ with $\Gamma_0 = 0.005$, where  
  the energy spectrum of the Dirac cone  $\xi_{\bk,\gamma,s}$ 
   is  valid for  $|\xi_{\bk,\gamma,s}| < 0.015$ from 
 Figs.~\ref{fig3} (b), (c) and (d). 
 Thus, Eq.~(\ref{eq:eq16a}) is valid   for a region  of $T < 0.01$, in which   
  the temperature corresponding to 
   the maximum of the conductivity  exists.

Figure \ref{fig6}(b) shows the  $T$ dependence of 
    $\sigma_{\pm}$, which is compared with  Fig.~\ref{fig5}(b).
A broad maximum is seen in  $\sigma_{+}$, and 
 the almost constant behavior of $\sigma_{-}$ is similar to 
that of Fig.~\ref{fig5}(b). 
  Compared with the inset of Fig.~\ref{fig5}(b), 
 a maximum of the angle $\phi$ in the inset of Fig.~\ref{fig6}(b) 
 is seen  due to a maximum in $\sigma_{xy}$. 
Such a maximum in the conductivity  can be understood    
    based on  
      a simplified model,~\cite{Suzumura_PRB_2018} where 
 $\sigma_+ \sim \sigma_- \sim \sigma$.
Note that 
 $\Gamma_{\rm ph}^\g$ is obtained in  Eq.~(\ref{eq:eq16a})
  and $\sigma  \simeq a_{\nu}'10^3T /\Gamma $ with  $a_{\nu}'= o(0.1)$
without the  e--p interaction. 
By taking $\Gamma$ replaced by $\Gamma + \Gamma_{\rm ph}^\g$ and 
employing an idea   $<|\xi_{\gamma,\bk}|> \sim T$ 
 with $<>$ being  an average value 
 in the summation of Eq.~(\ref{eq:self_energy}) (Appendix B), we obtain  
\begin{eqnarray}
\sigma \simeq \frac{a_{\nu}'10^3T}{1 + C_0RT^2/\Gamma} \; ,
  \label{eq:eq22b}
\end{eqnarray}
with $C_0$ = 50 and $\Gamma$ = 0.0005. 
Equation (\ref{eq:eq22b}) takes a maximum at $T= T_m = (\Gamma/C_0R)^{1/2}
 \simeq 0.003$, which is smaller than that of $\sigma_+$.
 Such $T_m$  can be improved 
 by noting  
 that  $T$ in the numerator of Eq.~(\ref{eq:eq22b}) may be 
  replaced by $T - \delta $ as seen from  Fig.~\ref{fig5}(b), e.g., 
  $T_m \simeq 0.06$ for $\delta = 0.002$.   
  From Eq.~(\ref{eq:eq22b}), it is found 
 that 
 a maximum of $\sigma$ as a function of $T$ is obtained  
 by a  competition of the  DOS (the numerator)
  and the e--p interaction (the denominator) 
 and that $\sigma$ decreases  with increasing $R$.

Figure \ref{fig6}(c) shows  $T$ dependence of 
 $1/\sigma_{\nu}$ ($\nu = x$ and $y$) for the transfer energy with 
 real  (dashed line) and 
 complex (solid line). 
$T$ independent behavior of $1/\sigma_{\nu}$ is seen 
at high temperatures due to the e--p interaction  and 
 the gradual increase of  $1/\sigma_{\nu}$ at low temperatures 
is similar to that of Fig.~\ref{fig5}(c).
The resistivity given by symbols, where  
 the difference between the solid line and symbols 
 are negligibly small,
 suggests the small effect of the e--p interaction 
 on the off-diagonal component $\sigma_{xy}$. 
 Since the effect of the e--p interaction is small for small $T$, 
 the insulating behavior at low temperature 
 comes from  the SOC. 

Finally, we examine  spin susceptibility, which is evaluated 
 from Eqs.~(\ref{eq:chi_a}) and  (\ref{eq:eq21}).
Figure  \ref{fig7} shows 
$T$ dependence of local magnetic susceptibility 
 $\chi_{\alpha}$ ($\alpha$ = A, A', B, and C), where 
 the dashed line (solid line) is calculated 
 for transfer energy with  real (complex).
The slight  difference between the dash and real lines suggests 
 that  the reduction of the magnetic susceptibility 
  by the SOC  is negligibly small 
    in contrast to the case of the conductivity. 
A relation $\chi_{A}=\chi_{A'}$ holds  due to 
 the inversion symmetry around the middle of A and A' sites. 
 Compared with that of $\alpha$-(ET)$_2$I$_3$ ,~\cite{Katayama_EPJ}  
  the susceptibility of $\chi_A (=\chi_A')$ and $\chi_C$ 
  shows 
   the rapid increase, 
 while the linear behavior  of $\chi_B$  is  a common feature.
 The total susceptibility of $\alpha$-(BETS)$_2$I$_3$ 
 ( $\chi^{\rm total}$) is shown by the dot-dashed line, which 
 is compared with that for $\alpha$-(ET)$_2$I$_3$.~\cite{Katayama_EPJ}  
The  noticeable increase for $\alpha$-(BETS)$_2$I$_3$ at finite temperatures 
is ascribed to a difference 
 in DOS close to the chemical potential as shown in Fig.~\ref{fig4}.
We note the low temperature behaviors of $\chi$ in Fig.~\ref{fig7}. 
For $\alpha$-(BETS)$_2$I$_3$, 
there is the following effect of the SOC on  $\chi$. 
Since the calculation was performed  only for $w_h,s=s'$, i.e., 
 the transfer energy of the  SOC with the same spin, 
 the insulating gap ($\simeq$ 0.001 eV) due to the opposite spin is absent.
However, 
 compared with $\chi$ of $\alpha$-(ET)$_2$I$_3$, which represents 
 $\chi_{\rm ET} \propto T$ in the absence of correlation, 
$\chi$ of $\alpha$-(BETS)$_2$I$_3$  shows  a slight reduction  
 from the  linear dependence 
and is convex downward for $T \sim $0.0005.   
  Although $\chi$ at lower temperatures  is not shown  due to 
  the numerical accuracy,
 it is expected that  $\chi$ reduces to zero linearly for $T \rightarrow 0$. 
Such a pseudogap behavior at low temperatures  comes from 
 the effect of the SOC with the same spin.

\section{Summary and Discussion}
We calculated the electric and magnetic properties of Dirac electrons 
in $\alpha$-(BETS)$_2$I$_3$ at ambient pressure and examined  
 the similarity and dissimilarity with those of $\alpha$-(ET)$_2$I$_3$ 
 at high pressures.~\cite{Suzumura2021_JPSJ}
 They show the  common feature of almost temperature independent conductivity 
 at high temperatures.  
The presence of the off-diagonal component ($\sigma_{xy}$), which is associated with both the tilting  of the Dirac cone and anisotropy of the velocity, 
results in the rotation of  the principal axis.
The crucial difference is the  SOC 
 in $\alpha$-(BETS)$_2$I$_3$, which gives rise to the reduction 
of the conductivity (or the enhancement  of the resistivity) 
 at low temperatures. 
We obtained the anisotropic conductivity with $\sigma_x > \sigma_y$ 
 due to the anisotropy of the velocity of the cone. 
 In contrast the opposite relation 
 $\sigma_y > \sigma_x$ is obtained 
 for  the previous case of the tilting along the $k_x$ -direction 
 with almost isotropic  velocity.~\cite{Suzumura2021_JPSJ} 
 The DOS exhibits a linear dependence around  the chemical potential. 
 Still,  such energy region is narrow compared with the previous case, 
~\cite{Suzumura2021_JPSJ} 
 since  $\eD$  is located close to the relevant  TRIM at M and Y points.
 Thus the temperature region for the linear susceptibility becomes narrow.

Here, we compare our result with that of the experiment.
The temperature dependence of resistance (corresponding to the inverse of the conductivity)  shows a nearly constant behavior at high temperatures 
 and noticeable increase at low temperatures. 
Our results are qualitatively consistent with those of the experiment 
 under ambient pressure.
\cite{Inokuchi1995_BCSJ68}
 The enhancement at low temperatures 
  comes from the interplay of the  effects of 
  the Dirac cone and  the SOC.
The SOC  has a significant effect on 
  the  diagonal transfer energy 
with both real and  imaginary parts. 
 The present result is a possible mechanism 
 for keeping an inversion symmetry between 
 A and A'. 

 Recent measurement of resistivity of $\alpha$-(BETS)$_2$I$_3$
\cite{Tajima2021PRB}  shows 
 an increase  at ambient pressure 
 but the almost constant behavior  under pressure $\sim$ 0.55 GPa 
 is quite similar to $\alpha$-(ET)$_2$I$_3$\cite{Tajima2007} 
suggesting Dirac fermion phase under pressure.
Thus, under pressure,  the effect of the SOC is reduced and the conventional Dirac cone 
 is expected due to short range repulsive interaction.\cite{Morinari2014}

We calculated the spin susceptibility 
 and found  the  rapid linear increase  at low temperatures 
compared with 
that of $\alpha$-(ET)$_2$I$_3$.\cite{Katayama_EPJ} 
 Such a linear increase  
  is compatible with a  measurement under ambient pressure
 of the susceptibility in $\alpha$-(BETS)$_2$I$_3$.~\cite{Fujiyama2021,Fujiyama2021_condmat}
Although the rapid decrease of the susceptibility is found 
 in $\alpha$-(ET)$_2$I$_3$ due to the  effect of the long range Coulomb interaction,
\cite{Hirata2016} 
 the linear behavior in $\alpha$-(BETS)$_2$I$_3$~\cite{Fujiyama2021} 
suggests that such a correlation effect is small 
 and the effect of the SOC is dominant for the insulating behavior.

\acknowledgements
We thank K. Yoshimi and M. Naka  
for valuable discussions.
This research was funded by a Grant-in-Aid for Scientific Research (19K21860) from the Japan Society for the Promotion of Science (JSPS) and JST, CREST Grant Number JPMJCR2094, Japan.
This work was performed under the GIMRT Program of the Institute for Materials Research (IMR), Tohoku University. 
TT is supported in part by the Leading Initiative for Excellent Young Researchers (LEADER), a program of the Ministry of Education, Culture, Sports, Science and Technology, Japan (MEXT). 
The DFT  computations were mainly conducted using the computer facilities of ITO at Kyushu University, MASAMUNE at IMR, Tohoku University, and ISSP, University of Tokyo, Japan. 

\appendix
\section{Matrix elements} 
 Using  first-principles calculations, 
 the TB model is obtained as~\cite{EPJB2020}
\begin{eqnarray}
H_0 = \sum_{i,j = 1}^N \sum_{\alpha, \sigma} \sum_{\beta, \sigma'}
 t_{i,j; \alpha,\beta; \sigma, \sigma'} 
 a^{\dagger}_{i,\alpha, \sigma} a_{j, \beta, \sigma'} \; , 
\label{eq:A1}
\end{eqnarray}
where 
  $t_{i,j; \alpha,\beta; \sigma, \sigma'}$ denotes 
 a  transfer energy obtained by  $(\bf{j}-\bf{i} = \bf{R})$ 
\begin{eqnarray}
t_{\alpha,\beta; \sigma, \sigma'}(\mathbf{R})=\langle\phi_{\alpha, \sigma,0}|H|\phi_{\beta, \sigma^{\prime},\mathbf{R}} \rangle .
\label{equ1}
\end{eqnarray}
The quantity  $\phi_{\alpha, \sigma,\mathbf{R}}$ is 
 the MLWF 
   spread over the molecule $\alpha$  and centered at $\mathbf{R}$. 
  Equation (\ref{equ1}) shows that 
 $t_{i,j; \alpha,\beta; \sigma, \sigma'}$ 
   depends only on the difference between the $i$-th site and the $j$-th site.

 We introduce  site-potentials acting on $B$ and $C$ sites, $\Delta V_{B}$ and $\Delta V_{C}$, which are measured from site-energy at $A$~($A^\prime$) site, $V_{A}$.\cite{Kondo2009}
\begin{eqnarray}
\Delta V_{B} = V_{B} - V_{A}, \\
\label{siteVb}
\Delta V_{C} = V_{C} - V_{A},
\label{siteVc}
\end{eqnarray}
where $V_{A}$, $V_{B}$, and $V_{C}$ are the site-energies at each molecule that  are calculated using MLWFs $|\phi_{\alpha,0} \rangle$; 
\begin{eqnarray}
V_{\alpha} =\langle\phi_{\alpha, \sigma,0}| H |\phi_{\alpha, \sigma^{\prime},0} \rangle, 
\label{eq:eqV33}
\end{eqnarray}
where ${\alpha}$ indicates $A$ (=~$A^{\prime}$), $B$, and $C$ molecules. 
These site-potentials are  listed in Table~\ref{Transfer_alpha}, where 
$\Delta V_{C}$  is modified from 0.0208 due to 
 a correlation effect.~\cite{EPJB2020}  
 In terms of  
    $X= \e^{i\kx}$, $\bar{X}=  \e^{-i\kx}$, 
    $Y= \e^{i\ky}$, and $\bar{Y}=  \e^{-i\ky}$, 
  matrix elements, $t_{ij} = (\hat{H})_{ij}$, are given by 
 \begin{eqnarray}
 t_{11} & = & t_{55} =   a_{1d}(Y+\bar{Y})
 +s1X+s1^*\bar{X} 
                          \; , \nonumber \\
 t_{22}& = & t_{66} =  a_{1d}(Y+\bar{Y})
 +s1^*X+s1\bar{X} 
                          \; , \nonumber \\
 t_{33}&=&  t_{77} =  a_{3d}(Y+\bar{Y})
     + s3(X + \bar{X})+ \Delta V_B
                           \; , \nonumber \\
 t_{44}&=&  t_{88} =  a_{4d}(Y+\bar{Y})
  + s4(X + \bar{X})+ \Delta V_C 
                           \; , \nonumber \\
 t_{12}&=&  t_{56} = a_3+a_2Y+d_0\bar{X}+ d_1XY 
                          \; , \nonumber \\
t_{13}&=&  t_{57} =  b_3+b_2\bar{X}+ c_2 \bar{X} Y + c_4\bar{X}\bar{Y}
                           \; , \nonumber \\
 t_{14}&=& t_{58} =  b_4Y+b_1\bar{X}Y+c_1\bar{X} + c_3
                           \; , \nonumber \\
t_{23} &=& t_{67} =  b_2+b_3\bar{X}+c_2 \bar{Y} + c_4 {Y}
                           \; , \nonumber \\
t_{24} &=&  t_{68} =  b_1+b_4\bar{X}+c_1Y + c_3\bar{X}Y
                           \; , \nonumber \\
t_{34}&=&  t_{78} = a_1+a_1Y + d_2\bar{X}+d_3X 
                   + d_2 XY+d_3\bar{X}Y 
                               \; ,   \nonumber \\
t_{17} &=& b2_{so1} \bar{X} + c2_{so1}\bar{X} Y
                           + c4_{so1}\bar{X} \bar{Y}
                           \; , \nonumber \\
t_{18} &=&   b1_{so1}\bX Y  + b4_{so1} Y + c1_{so1} \bX 
                           \; , \nonumber \\
t_{27} &=& b2_{so1}  + c2_{so1} \bY + c4_{so1} Y 
                           \; , \nonumber \\
t_{28} &=& b1_{so1}  + c1_{so1}Y + c3_{so1}\bX Y 
                           \; , \nonumber \\
t_{35} &=& b2_{so2} X  + c2_{so2}X \bY + c4_{so2}X Y
                           \; , \nonumber \\
t_{36} &=& b2_{so2}  + c2_{so2} Y  + c4_{so2} \bY 
                           \; , \nonumber \\
t_{45} &=&  b1_{so2}X \bY  + b4_{so2}\bY  + c1_{so2}X
                           \; , \nonumber \\
t_{46} &=& b1_{so2}  + c1_{so2} \bY + c3_{so2}X  \bY 
                           \; ,  
\label{eq:matrix_element}
\end{eqnarray}
 $t_{15} = t_{16} = t_{25} = t_{26}
 =t_{37} = t_{38} = t_{47} = t_{48} = 0$, and 
 $t_{ji} = t_{ij}^*$.

\section{Damping by phonon scattering}

For the electric transport, we calculate dampings of 
impurity and phonon scattering. 
In Eq.~(\ref{eq:H}), the third term denotes  
  the harmonic phonon   given by  
 $H_{\rm p}= \sum_{\bq} \omega_{\bq} b_{\bq}^{\dagger} b_{\bq}$ 
 with $\omega_{\bq} = v_s |\bq|$ and  $\hbar$ =1 ,and
 the fourth term is  the e--p interaction  expressed 
~\cite{Suzumura_PRB_2018}
\begin{equation}
 H_{\rm e-p} = \sum_{\g,s} \sum_{\bk} \sum_{\bq}
   \alpha_{\bq} c_{\g s}(\bk + \bq)^\dagger c_{\g s}(\bk) \phi_{\bq} \; ,
\label{eq:H_e--p}
\end{equation}
 with 
 $\phi_{\bq} = b_{\bq} + b_{-\bq}^{\dagger}$.
  We introduce  a coupling constant $\lambda = |\alpha_{\bq}|^2/\omega_{\bq}$, 
  which becomes  independent of $|\bq|$  for small $|\bq|$. 
The e--p scattering is considered 
 within  the same band (i.e., intraband) 
 owing  to the energy conservation with $v \gg v_s$, where 
  $v \simeq 0.05$~\cite{Katayama_EPJ} 
 denotes the averaged velocity of the Dirac cone. 
The last term of Eq.~(\ref{eq:H}), $H_{\rm imp}$, denotes a normal  impurity 
 scattering, which gives   
 a constant conductivity.

The damping of electrons of the $\g$ band, which is defined by 
 $\Gamma_\g$,  is obtained from the electron Green function\cite{Abrikosov} 
 expressed as   
\begin{subequations}
\begin{eqnarray}
 G_\g(\bk, i \omega_n)^{-1} & = & 
 i \omega_n - E_{\g,\bk}+ \mu 
  + i \Gamma_{\g}  
  \; ,
 \label{eq:eq14a} \\
\Gamma_{\g} & = & \Gamma + \Gamma_{\rm ph}^{\g}
 \; , 
 \label{eq:damping} 
  \end{eqnarray} 
\end{subequations}
 where 
$\Gamma_{\rm ph}^{\g} = - {\rm Im} \Sigma_\g (\bk, E_{\g, \bk} - \mu)$ 
 with $ \Sigma_\g (\bk, E_{\g, \bk} - \mu)$
 being  a self-energy given by the e--p interaction. 
 The real part of the self-energy 
 can be neglected for  doping at low concentrations.~\cite{Suzumura_PRB_2018}
 The quantity $\Gamma$ comes from another  self-energy by 
the impurity scattering.
Note that $\Gamma_{\rm ph}^\g$ does not depend on $\Gamma$,
 and that the ratio $\Gamma_{\rm ph}^{\g}/\Gamma$ 
  is crucial to the determination of  
 the $T$ dependence of the conductivity.    
The quantity $\Sigma_\g (\bk, \omega) = \Sigma_\g (\bk, i \omega_n)$ 
 with $i\omega_n \rightarrow \omega +  0$ 
 is  estimated from  
\cite{Abrikosov} 
\begin{eqnarray}
 & & \Sigma_\g (\bk, i \omega_n)  =  T \sum_m \sum_{\bq}\; |\alpha_q|^2 
            \nonumber \\
 & &\times   \frac{1}{i \omega_{n+m} - \xi_{\g, \bk+\bq}} 
      \times \frac{2 \omega_{\bq}}{\omega_{m}^2 + \omega_{\bq}^2} \; , 
 \label{eq:self_energy}
  \end{eqnarray} 
which is a product of electron and phonon Green functions. 
$\omega_n=  (2n+1)\pi T$, $\omega_{m}=2\pi m T$ with $n$ and $m$ being integers. $\xi_{\g, \bk} = E_{\g, \bk} - \mu$. 
 Applying the previous result,\cite{Suzumura_PRB_2018}
 we obtain Eqs.~(\ref{eq:eq16a}) and (\ref{eq:eq16b}).


\end{document}